\documentstyle[preprint,aps]{revtex}
\tightenlines
\def\L{\Lambda}
\def\bea{\begin{equation}}
\def\ena{\end{equation}}
\def\bey{\begin{eqnarray}}
\def\beyn{\begin{eqnarray*}}
\def\eny{\end{eqnarray}}
\def\enyn{\end{eqnarray*}}

\begin{document}
\begin{titlepage}
\begin{flushright}
IFUM 614/FT\\
March 1998 \\
\end{flushright}
\vspace{1.5cm}
\begin{center}
{\bf \large WILSONIAN FLOW AND MASS-INDEPENDENT RENORMALIZATION }
\footnote{Work supported in part by M.U.R.S.T.}\\
\vspace{1 cm}
{ M. PERNICI} \\ 
\vspace{2mm}
{\em INFN, Sezione di Milano, Via Celoria 16, I-20133 Milano, Italy}\\
\vspace{0.6 cm}
{ M. RACITI }\\
\vspace{2mm}
{\em Dipartimento di Fisica dell'Universit\`a di Milano, I-20133 Milano,
Italy}\\
{\em INFN, Sezione di Milano, Via Celoria 16, I-20133 Milano, Italy}\\
\vspace{2cm}
\bf{ABSTRACT}
\end{center}
\begin{quote}
  
~~~We derive the Gell-Mann and Low renormalization group equation 
in the Wilsonian approach to renormalization of massless $g\phi^4$ 
in four dimensions, as a particular case of a non-linear equation
satisfied at any scale by the Wilsonian effective action.

We give an exact expression for the $\beta$ and $\gamma_{\phi}$
functions in terms of the Wilsonian effective action 
at the Wilsonian renormalization scale $\L_R$; at the first two loops 
they are simply related to the gradient of the flow of
the relevant couplings and have the standard values; beyond two loops
this relation is spoilt by corrections due to irrelevant couplings.

We generalize this analysis to the case of massive $g\phi^4$,  
introducing a mass-independent Wilsonian renormalization scheme;
using the flow equation technique we prove renormalizability 
and we show that the limit of vanishing mass parameter exists.
We derive the corresponding renormalization group equation, in which 
$\beta$ and $\gamma_{\phi}$ are the same as in the massless case;
$\gamma_m$ is also mass-independent; at one loop it is the gradient
of a relevant coupling and it has the expected value.

\end{quote}
\end{titlepage}
\section*{Introduction}

The renormalization group equation describes the response of the
parameters in the renormalization conditions due to a change in
the renormalization scale \cite{Rm}. It provides useful information on
the asymptotic behavior of the Green functions at large momenta.

A standard way of introducing the renormalization scale is to impose
the renormalization conditions at a momentum subtraction point \cite{Rm,Sy}.

In the Wilsonian renormalization group approach \cite{Wil} the 
renormalization scale $\L_R$
separates the hard modes  with $p > \L_R$ from the soft
modes with $p < \L_R$; roughly speaking,
only hard modes propagate in the internal
lines of the graphs of the Wilsonian effective action. The exact way in
which the soft modes are frozen depends on the form of the cut-off
function. 

In the hard-soft (HS) renormalization schemes, first introduced in
\cite{LM} with the purpose of studying in a simple way the
renormalizability of massless theories with BPHZ, a splitting in hard
and soft fields is made at a scale $\L_R$, in such a way that the 
renormalization conditions can be chosen at zero momentum even 
in massless theories. The independence of the physical 
quantities from such a splitting is guaranteed by the renormalization 
group equation obtained using the Quantum Action Principle \cite{Qap}.

A recent discussion of the HS schemes in the Wilsonian approach
can be found in \cite{PRR}.
The problem of deducing the renormalization group equation within
these Wilsonian HS schemes has not yet been solved for a generic
HS cut-off function.

In the case of step-function cut-off, the renormalization
group equation has been obtained in $g\phi^4$ in \cite{wein}; 
the beta function and $\gamma_{\phi}$ are simply related to the
gradient of the flow of the relevant couplings.
It is an interesting question whether this property holds for smoother
cut-off functions, which are more suitable in perturbative Quantum
Field Theory.

Using the flow equation \cite{Wil,Polchi}
in the case of smooth cut-off with compact support, an approximate
derivation of the renormalization group equation has been given in
massive
$g\phi^4$ 
in a HS scheme in \cite{HL}. Their formulae for the beta and gamma
functions have the above-mentioned relation to the flow of the
relevant couplings.
 They verified this gradient flow relation explicitely at the first 
non-trivial loop order for the beta function and 
 $\gamma_{\phi}$, showing that in the massless
limit they have the standard values, which are expected to be 
scheme-independent \cite{Gross}.

The main results of this paper are the following: 

i) we introduce
an  `effective  renormalization group equation' for the Wilsonian
effective action, in terms of which
we obtain in the Wilsonian HS renormalization schemes 
exact formulae (in perturbation theory) for the beta and the
gamma functions in terms of the Wilsonian effective action at scale $\L_R$;

ii) we introduce a mass-independent HS scheme for massive $g\phi^4$,
which admits zero mass limit; we deduce the corresponding
mass-independent renormalization group equation.
These results are first discussed heuristically in a general setting,
then are proven in the framework of the flow equation technique
\cite{Polchi}.

We start observing that, assuming the validity of the
 renormalization group equation,
 it is possible to obtain its extension
to the case of the Wilsonian effective action at scale $\L > 0$.
This is an effective
renormalization group equation which is non-linear; in the limit of
$\L \to 0$ it becomes the usual renormalization group equation.
 The situation is analogous to the Ward identities in gauge theory;
at scale $\L \neq 0$ the Wilsonian effective action satisfies `effective
Ward identities' \cite{Becchi2}, which in the limit $\L \to 0$ become
the usual Ward identities.
In both cases there are non-linear terms due to the fact that, in
 general, field transformations (respectively rescalings and gauge
 transformations) do not commute with the Wilsonian renormalization flow.

In the massive case we study in detail 
the mass-independent renormalization group equation, first introduced
by Weinberg \cite{wein2}. This equation has the advantage, in
comparison with the Gell-Mann and Low \cite{Rm} or 
the Callan-Symanzik \cite{CS}
equations, that it can be solved exactly; this is a useful property in
studying the asymptotic behavior of the Green functions.
The HS schemes have, by definition, renormalization conditions
 at zero momentum even in massless theories; 
it is a natural step to impose renormalization conditions at zero
 momentum and zero mass in a generic theory.
We introduce a mass-independent HS renormalization scheme in massive
$g\phi^4$, which is similar in spirit
to the mass-independent scheme
introduced by Weinberg \cite{wein2}; in the latter case there
are problems in dealing with scalars, due to the fact that mass insertions on
the massless theory are infrared-singular.
In the HS schemes these infrared singularities are absent; 
in fact mass insertions are made on the Wilsonian vertices at $\L_R$,
which are trivially infrared finite.
The HS mass-independent renormalization scheme described here is also
related to the mass-independent scheme in \cite{GS}, which however applies
specifically to massive theories, while the HS schemes apply equally
well to massless theories which cannot be considered the massless
limit of a massive theory, as originally remarked in \cite{LM}.
Using the effective renormalization group equation 
at scale $\L_R$, we obtain in these schemes 
exact formulae (in perturbation theory) for the beta and the
gamma functions in terms of
the Wilsonian effective action at scale $\L_R$.
$\beta$ and  $\gamma_{\phi}$ are the same as in the
massless case.
At the first two loops they are simply
related to the gradient of the flow of the relevant couplings. 
As a check, we verify by explicit computation that for a class of
analytic cut-off functions they have the standard values, which is
 expected to be scheme independent \cite{Gross}.
$\gamma_m$ is also mass-independent; at one loop it has the
gradient flow expression and  the standard value.
Beyond two loops  this gradient flow property fails for generic
 cut-off functions.

In the particular case in which the cut-off function characterizing
the HS scheme is smooth and with compact support, 
these results can be put on a rigorous footing using the flow equation
technique \cite{Polchi}.
First of all we prove the renormalization group equation for massless
$g\phi^4$ ; this proof involves a version of the Quantum Action
Principle \cite{Qap}, discussed with the flow equation technique 
in \cite{Becchi2}.
Then we treat the massive case in a similar way, using our
mass-independent HS renormalization scheme.
Using the flow equation we prove the ultraviolet and
infrared convergence of the theory using appropriate bounds. 
We prove the effective mass-independent renormalization group
equation and we show that for $\L \to 0$ it becomes the usual
mass-independent renormalization group equation.

In the first section we introduce the mass-independent HS schemes;
 we obtain the effective renormalization group
 equation for the Wilsonian effective action, leading to 
exact formulae for the beta and gamma functions in terms of the Wilsonian 
vertices at scale $\L_R$; we compute them at low loop orders.
In the second section we prove, using the flow equation,
the renormalization group equation for the massless $g\phi^4$ theory.
In the third section and in the Appendix we prove the consistency of
the  mass-independent HS schemes and we
derive the renormalization group equation  for massive $g\phi^4$.

\section{Effective renormalization group equation
and mass-independent renormalization}

\subsection{Effective renormalization group equation}

Consider the $g\phi^4$ theory in Euclidean four-dimensional
space; the path-integral is 

\bea\label{Z}
Z_{0\L_0}[J] \equiv  
 \int{\cal D}\Phi e^{- {1\over \hbar} [S(\Phi) - J \Phi]} 
\ena
 with the bare action
\bey\label{barem}
&&S (\Phi)= {1\over2}\Phi D_{0\L_0}^{-1}\Phi + S^I(\Phi) \\
&&S^I(\Phi) = {1\over 2} \int_p \phi (-p)[ c_1 +
c_2p^2] \phi (p)
+ {c_3 \over 4!} \int_{p_1p_2p_3} \phi (p_1) \phi (p_2)\phi (p_3)
\phi (p_4)  \nonumber
\eny
where  $p_4=-p_1-p_2-p_3$ and
$S^I$ is the interacting part of the bare action. 
We use a compact notation in which $\Phi =\{ \phi (p) \}$ and
 a matrix notation is used for integrations
on the momenta; furthermore we will use the trace operation meaning 
integration over momentum $\int_p \equiv \int {d^4p \over (2\pi)^4}$.

$D_{0\L_0}(m,p)$ is the propagator with
ultraviolet cut-off $\L_0$, which for $\L_0 \to \infty$ converges to
$D(m,p)={1\over p^2+m^2}$.
For the moment we will not specify the ultraviolet regulator used,
since our argument will not depend on it.
Let us split the propagator in two parts characterized by a scale $\L > 0$
\bey
D_{0\L_0}= D_S + D_H ~~;~~ D_H = D_{\L\L_0}
\eny
where the `hard' propagator $D_H$ behaves as $D_{0\L_0}$ for large
momenta, while it is regular in $p=0$ in the massless case, 
~$\L$ acting as an infrared
cut-off on this propagator; the `soft' propagator $D_S$ converges
fast enough to zero for large momenta, and it behaves as $D_{0\L_0}$
for small momenta in the massless case.
Consider a cut-off function 
$K_{\Lambda}(m,p)$ satisfying $K_{\Lambda}(0,0)=1$ and going to zero 
at least as $\L^4/(p^2)^2$ for $p^2/\L^2 \to \infty$.
We require that
\bey
D_H = D_{\L\L_0} \to D(1-K_{\L})
\eny
 for $\L_0 \to \infty$.
For the considerations in this section this function is understood to be 
analytic. 

Let us make an `incomplete integration'  over the hard
modes 
\bea\label{Zj}
Z_{\L\L_0}[J] \equiv  
 \int{\cal D}\Phi 
e^{-{1\over \hbar}[{1\over2}\Phi D_{\L\L_0}^{-1}\Phi+ S^I(\Phi) - J \Phi]}  
\ena
The flow of this functional from $\L$ to zero can be represented as
\bea\label{ZL}
Z_{0\L_0}[J] =  
e^{{\hbar \over 2} {\delta \over\delta J} (D_{\L\L_0}^{-1}- D_{0\L_0}^{-1})
{\delta \over\delta J}} Z_{\L\L_0}[J]
\ena
$Z_{\L\L_0}[J]$ is infrared finite for $\L > 0$ even in the massless
limit. In all the formulae involving functionals used in this paper,
only the source ( or field) dependent terms are well-defined.

The $1$-PI functional generator corresponding to $Z_{\L\L_0}$ is
\bea\label{GL}
\Gamma_{\L\L_0}[\Phi] = {1\over2}\Phi D_{\L\L_0}^{-1}\Phi +
\Gamma^I_{\L\L_0}[\Phi]
\ena
The $n$-point $1$-PI effective Green function
$\Gamma_n^{\L\L_0}(m;p_1,...,p_{n-1})$ depends on the mass $m$ and on
$n-1$ independent momenta.
 
Let us introduce a renormalization scheme in which some
renormalization scale $\L_R$ appears; according to general arguments 
the Gell-Mann and Low renormalization group equation 
on $Z[J]\equiv lim_{\L_0 \to \infty}Z_{0\L_0}[J]$  holds 
\bea\label{rge}
(\L_R {\partial \over \partial \L_R }+
\beta {\partial \over \partial g } + 
\gamma_{\phi} J{\delta \over \delta J } 
- \gamma_{m} m^2{\partial \over \partial m^2 } ) Z[J] = 0
\ena
where $\beta, \gamma_{\phi}$ and $\gamma_m$ are in general functions
of $g$ and ${m \over \L_R}$. 
From eq.(\ref{ZL}) it follows that 
\bea
Z_{\L}[J]= e^{{1\over \hbar}W_{\L}[J]} \equiv Z_{\L\infty}[J] 
\ena
satisfies the `effective renormalization group equation'
\bea\label{rgef}
(\L_R {\partial \over \partial \L_R }+ 
\beta {\partial \over \partial g } + 
\gamma_{\phi} J{\delta \over \delta J }
- \gamma_{m} m^2{\partial \over \partial m^2 }
-\hbar{\delta \over\delta J} A_{\L} {\delta \over\delta J} ) Z_{\L}[J] = 0
\ena
where
\bea\label{A}
A_{\L} \equiv (\gamma_{\phi} + {1\over 2} \gamma_{m} 
m^2{\partial \over \partial m^2 }) ( {D^{-1} K_{\L} \over 1-K_{\L}})
\ena
In terms of $W_{\L}$ the effective renormalization group equation reads
\bea
(\L_R {\partial \over \partial \L_R }+ 
\beta {\partial \over \partial g } + 
\gamma_{\phi} J{\delta \over \delta J }
- \gamma_{m} m^2{\partial \over \partial m^2 }) W_{\L}
-{\delta W_{\L} \over\delta J} A_{\L} {\delta W_{\L}  \over\delta J} 
- \hbar tr A_{\L} {\delta^2 W_{\L}  \over\delta J^2}  = 0
\ena
Observe that the source field rescaling operator $J{\delta \over \delta J }$
and the mass rescaling operator $m^2{\partial \over \partial m^2 }$
do not commute with the flow, leading to the last two terms in the
above equation; analogous terms appear in the effective gauge Ward identity
\cite{Becchi2}.

Making a Legendre transformation we get
\bea
(\L_R {\partial \over \partial \L_R }+ 
\beta {\partial \over \partial g } - 
\gamma_{\phi} \Phi{\delta \over \delta \Phi }
- \gamma_{m} m^2{\partial \over \partial m^2 }) \Gamma_{\L} + \Phi A_{\L} \Phi 
+\hbar tr A_{\L} ({\delta^2 \Gamma_{\L}  \over\delta \Phi^2})^{-1}  = 0
\ena
Using (\ref{GL}) we get the effective renormalization group 
equation on $\Gamma_{\L}^I[\Phi] \equiv \Gamma_{\L\infty}^I[\Phi]$ in the form
\bey\label{rgefG}
&&(\L_R {\partial \over \partial \L_R }+ 
\beta {\partial \over \partial g } - 
\gamma_{\phi} \Phi{\delta \over \delta \Phi }
- \gamma_{m} m^2{\partial \over \partial m^2}) \Gamma^{I}_{\L}[\Phi] 
= \nonumber \\
&&= \Phi (\gamma_{\phi} D^{-1} +{m^2 \over 2} \gamma_{m} ) \Phi
-\gamma_{\phi} {\cal T}_{\phi}^{\L}[\Phi]-\gamma_{m}{\cal T}_{m}^{\L}[\Phi] 
\eny
with the following non-linear terms
\bey\label{T}
{\cal T}_{\phi}^{\L}[\Phi] \equiv
\hbar tr  ( {D^{-1} K_{\L} \over 1-K_{\L}})
 ({\delta^2 \Gamma_{\L}  \over\delta \Phi^2})^{-1}~~;~~
{\cal T}_{m}^{\L}[\Phi]  \equiv
{\hbar \over 2} tr [ m^2{\partial \over \partial m^2} 
( {D^{-1} K_{\L} \over 1-K_{\L}})]
 ({\delta^2 \Gamma_{\L}  \over\delta \Phi^2})^{-1}
\eny
The $n$-point functions
${\cal T}_n^{\L}(m;p_1,...,p_{n-1})$ are defined factorizing as usual
the delta function for momentum conservation.
From (\ref{rgefG}) one obtains the effective renormalization group
equations on $\Gamma_n^{\L}(m;p_1,...,p_{n-1})$; we do not give any
meaning to the field independent part of (\ref{rgefG}).

Imposing  Wilsonian renormalization conditions at $\L = \L_R > 0$, one can
use the effective renormalization group equation (\ref{rgefG})
to give an expression of the  beta and gamma functions in terms of the
Wilsonian effective action.
A standard set of zero-momentum renormalization conditions at a
Wilsonian scale $\L_R > 0$  for the massive $g\phi^4$ theory is
the following \cite{Polchi,HL}:
\bey\label{renG3}
\Gamma^{I \L_R\L_0(l)}_2(m;0) = 0~~;~~
\partial_{p^2}|_{p=0} \Gamma^{I \L_R\L_0(l)}_2(m;p) =0~~;~~
\Gamma^{I \L_R\L_0(l)}_4(m;0,0,0) =g~ \delta^{l,0}
\eny 
where $l$ is the loop index in the perturbative expansion.
In the following we will call for short HS (hard-soft) scheme any
renormalization scheme in which zero-momentum renormalization
conditions at a Wilsonian scale $\L_R > 0$ are used. For a discussion
of these schemes see \cite{PRR,El}.

For the beta and gamma functions one finds, using (\ref{rgefG},\ref{renG3})
\bey\label{BETAA}
\beta =\beta_3 +\gamma_{\phi} [4g +\alpha_3 ]+\gamma_m \alpha_4~~;~~
\gamma_{\phi}= -{\beta_2 +\gamma_m \alpha_5 \over 2+\alpha_2}~~;~~
\gamma_m =-{ \beta_1 + \gamma_{\phi}(2+\alpha_1)\over 1+\alpha_6}
\eny
where 
\bey
\beta_1 = \L {\partial \over \partial \L } |_{\L_R} m^{-2}
\Gamma^{I~\L}_2(m;0)~~;~~
\beta_2 = \L {\partial \over \partial \L } |_{\L_R} 
\partial_{p^2}|_{0} \Gamma^{I~\L}_2(m;p) ~~;~~
 \beta_3 = \L {\partial \over \partial \L } |_{\L_R}
\Gamma^{\L}_4|_{p_i=0}
\eny
are the gradients of the flow of the relevant terms at $\L_R$, and the
alpha coefficients are related to the non-linear terms of (\ref{rgefG})
\bey
\alpha_1 =&&-m^{-2}{\cal T}_{\phi~2}^{\L_R}(m;0)~~;~~
\alpha_2 =- \partial_{p^2}|_{0} {\cal T}_{\phi~2}^{\L_R}(m;p)~~;~~
\alpha_3 = - {\cal T}_{\phi~4}^{\L_R}(m;0,0,0)  \nonumber \\
\alpha_4 =&& -  {\cal T}_{m~4}^{\L_R}(m;0,0,0)~~;~~
\alpha_5 = - \partial_{p^2}|_{0} {\cal T}_{m~2}^{\L_R}(m;p)~~;~~
\alpha_6 = - m^{-2} {\cal T}_{m~2}^{\L_R}(m;0)
\eny
Although for semplicity we have derived the above equations in the
case of analytic cut-off, it is straightforward to generalize them to
the case of smooth functions $K_{\Lambda}(m,p)$ with compact support, 
which we will use in the following sections.

In the limit of step-function cut-off $K_{\L}(m;p) \to \theta (\L - |p|)$
the $n$-point functions ${\cal T}_n^{\L}(m;p_1,...,p_{n-1})$ vanish
since
$K_{\L}(1-K_{\L}) \to 0$, so that the Wilsonian vertices satisfy in
this limit the ordinary renormalization group equation;
the beta and the gamma functions
are gradients of the flow of relevant couplings, since their alpha
terms vanish, in agreement with  \cite{wein}.

For a generic cut-off function  the alpha  terms in the beta and
the gamma functions are absent
at lowest orders in perturbation theory, so that the gradient flow
property holds in this approximation, as checked in \cite{HL}; however
this property fails at higher loops.

To our knowledge, exact relations (in perturbation theory) for the beta
and gamma functions in terms of the Wilsonian effective action at the
renormalization scale have not been given before; for instance in
\cite{HL} the alpha terms in (\ref{BETAA}) are absent.

\subsection{Mass-independent HS renormalization scheme} 

In the following we will restrict our attention to mass-independent 
HS renormalization schemes which admit a straightforward massless limit.
The reason of interest of mass-independent renormalization schemes, 
as pointed out by Weinberg 
\cite{wein2} is that the beta and gamma functions are
mass-independent, allowing a simple solution to the renormalization
group equation (\ref{rge}).

In these renormalization schemes the bare action (\ref{barem}) depends
only polynomially on the mass; in particular $c_2$ and $c_3$ are
mass-independent, while the mass counterterm has the following form
\bey\label{MMM}
c_1 = \hat{c}_1\L_0^2 + \tilde{c}_1 m^2
\eny
where $\hat{c}_1$ and $ \tilde{c}_1$ are mass-independent.

Let us introduce the following mass-independent zero-momentum
renormalization conditions at a renormalization scale $\L_R > 0$ :
\bey\label{renG0}
\Gamma^{I \L=0\L_0(l)}_2(0;0) = 0 
\eny
\bey\label{renG1}
\partial_{p^2}|_{p=0} \Gamma^{I \L_R\L_0(l)}_2(0;p) =0~~;~~
\Gamma^{I \L_R\L_0(l)}_4(0;0,0,0) =g~ \delta^{l,0}
\eny 
and
\bey\label{renG2}
\partial_{m^2}|_{m=0} \Gamma^{I \L_R\L_0(l)}_2(m;0) =0
\eny
In (\ref{renG0}-\ref{renG2}) there is one more condition than in 
 (\ref{renG3}); on the other hand in (\ref{MMM}) an extra bare
parameter appears, so that the number of bare parameters matches again
with the number of renormalization conditions;
 the fourth renormalization condition (\ref{renG2})
fixes the extra parameter $ \tilde{c}_1$.

We will prove later (see Appendix B) that the theory defined by  
(\ref{renG0}-\ref{renG2})
becomes, for $m \to 0$, the massless theory associated with the
conditions  (\ref{renG0},\ref{renG1}) only.

The condition (\ref{renG0}) is the one which is necessary to impose
in  massless theories; it guarantees that a self-energy subdiagram insertion
provides the $p^2$ factor needed to cancel the extra $1/p^2$ propagator. 
In the conditions (\ref{renG1})
the scale $\L_R$ plays a role similar to the non-zero momentum
subtraction point in a massless theory. The condition (\ref{renG2})
 has no equivalent at $\L = 0$ and non-zero momentum 
subtraction point; in fact
$\partial_{m^2}\Gamma^{I \L=0 \L_0(l)}_2(m;p)|_{p^2=\mu^2} $ does not
exist at $m=0$, since mass insertions on the 
internal lines of a massless graph diverge in the
infrared region of loop momenta for any value of the external momenta.
 Obviously for $\L_R > 0$ these divergences are
absent. In theories without scalars,  analogous renormalization
conditions at $\L = 0$ and  non-zero momentum subtraction point
 can be imposed \cite{wein2}; e.g. on the fermionic
two-point function in QED one can choose the condition
$\partial_{m}|_{m=0} \Gamma^{I \L=0 \L_0(l)}_2(m;p)|_{p^2=\mu^2} =0$.

Let us discuss in a qualitative way the consistency of this HS
scheme; it is useful to treat separately the ultraviolet and the
infrared problems. 
To show that it leads to ultraviolet renormalization, observe
that substituting the condition (\ref{renG0}) with
\bea\label{reno}
\Gamma^{I \L_R\L_0(l)}_2(0;0) = \rho_1^{(l)}
\ena
(with $\rho_1^{(0)}=0$ ) 
and keeping the conditions  (\ref{renG1},\ref{renG2}), 
the Wilsonian vertices $\Gamma^{\L_R\L_0}_n$, constructed with the
bare action (\ref{barem}) and the scalar propagator $D_{\L_R\L_0}$,
are renormalized. The condition (\ref{renG2}) can be consistently
imposed since $\L_R$ acts as an infrared cut-off on the propagator.
This condition is a satisfactory renormalization condition for the
massive theory since in this scheme the derivatives with respect to
$m^2$, applied to a single graph in which subdivergences have been
removed, lower by two the degree of divergence. In a scheme which is
not mass-independent this fact would not be true, since the mass
derivative can act on the factors $ln m^2$ in the counterterms.
The theory at
$\L=0$ can be constructed with these Wilsonian vertices and soft
propagators $D_{0\L_R}$.
Since the  Wilsonian vertices are renormalized and the
soft propagators go to zero sufficiently fast (at least as 
$\L_R^4/(p^2)^3$)
at large momentum, by power counting it follows that the graphs
are superficially convergent, so that by the usual arguments one
expects that at $\L=0$ the theory is ultraviolet finite
(a proof of this fact will be given in the case of
compact-support cut-off in Appendix A).
For $m \neq 0$ there are no infrared problems in this procedure; however for
generic values of $\rho_1$ in  (\ref{reno}) the quantity
  $\Gamma^{I \L=0\L_0(l)}_2(m;0)$ will not admit limit for 
$m \to 0$, so that the  condition (\ref{renG0}) is not
satisfied, and the massless limit of the theory does not exist.

It is interesting to notice that,
choosing $\rho_1^{(l)} = \L_R^2 f^{(l)}(g;{\L_R \over \L_0})$ 
in such a way that (\ref{renG0}) holds order by order in loops,
 the limit $m \to 0$ can be made on the
Green functions with non-exceptional momenta, obtaining the 
corresponding Green functions of the massless theory
 (a proof of this fact will be given in the case of
compact-support cut-off in Appendix B).

In this HS scheme the bare coefficients $\hat{c}_1, \tilde{c}_1, c_2$ and
$c_3$ in (\ref{barem},\ref{MMM}), determined loop by loop in terms of the
renormalization conditions (\ref{renG0}-\ref{renG2}), are mass-independent. 

Using the renormalization conditions (\ref{renG1},\ref{renG2}) in equation
(\ref{rgefG}) at $\L = \L_R$ we get the following exact expressions for
the coefficient functions of the renormalization group equation in
terms of the Wilsonian action:
\bey\label{beta}
\gamma_{\phi} = {-\beta_2 \over 2+\alpha_2}~~;~~
\beta = \beta_3 + \gamma_{\phi} (4g + \alpha_3)~~;~~
\gamma_m = - {\beta_1+\gamma_{\phi}(2+\alpha_1) \over 1+\alpha_m} 
\eny
where now the gradients of the flow of the relevant couplings are
\bey\label{Wbeta}
\beta_1 =  \L {\partial \over \partial \L } |_{\L_R}
\partial_{m^2} \Gamma^{I~\L}_2|_0~~;~~
\beta_2 = \L {\partial \over \partial \L } |_{\L_R} 
\partial_{p^2} \Gamma^{I~\L}_2|_0~~;~~
\beta_3 = \L {\partial \over \partial \L } |_{\L_R}
\Gamma^{\L}_4|_0
\eny
and $(..)|_0$ indicates evaluation at zero mass and momenta;
the alpha coefficients 
\bey\label{a2}
\alpha_1 \equiv  -\partial_{m^2} {\cal
T}_{\phi~2}^{\L_R}|_0~~;~~
\alpha_m \equiv - \partial_{m^2}{\cal T}_{m~2}^{\L_R}|_0~~;~~
\alpha_2 \equiv - \partial_{p^2} {\cal T}_{\phi~2}^{\L_R}|_0~~;~~
\alpha_3 \equiv  -  {\cal T}_{\phi~4}^{\L_R}|_0
\eny 
represent the correction due to the non-linear terms in the
effective renormalization group equation (\ref{rgefG}); they are
vanishing at tree level.

\subsection{Low order computations}

According to general arguments \cite{Gross} one expects that
$\gamma_m^{(1)}, \beta^{(1)},\gamma_{\phi}^{(2)}$ and $ \beta^{(2)}$
are the same in all mass-independent renormalization schemes 
 (while $\gamma_{\phi}^{(1)}$ is trivially zero).
We will verify this explicitly for a class of HS schemes.
Observe that from (\ref{beta}), $\gamma_{\phi}^{(1)}=0$ and the
vanishing at tree level of the alpha coefficients in (\ref{a2}),  
it follows that 
$\gamma_m^{(1)}, \beta^{(1)},\gamma_{\phi}^{(2)}$ and $ \beta^{(2)}$
depend in a simple way on the gradient of the flow of the relevant 
couplings at $\L_R$, in agreement with analogous formulae in \cite{HL}.

$\beta^{(1)}$ and $\gamma^{(2)}_{\phi}$ have been computed in \cite{HL} in a
HS scheme using a
generic smooth compact-support cut-off, giving the standard value in
the massless limit. Let us compute $\gamma_m^{(1)}$,
$\beta^{(1)}$, $\beta^{(2)}$ and 
$\gamma^{(2)}_{\phi}$ in a class of mass-independent HS schemes with 
analytic cut-off.

Using the renormalization conditions
(\ref{renG0}-\ref{renG2})
the non-vanishing one-loop bare parameters are the following
\bey
\hat{c}_1^{(1)} = {-g \over 2\L_0^2} \int_q D_{0\L_0}(q)~~;~~
\tilde{c}_1^{(1)} = -{g \over 2} \int_q 
\partial_{m^2}|_0D_{\L_R\L_0}(m;q)~~;~~
c_3^{(1)} = {3g^2 \over 2} \int_q D_{\L_R\L_0}^2(q)
\eny
where we define $D_{\L\L_0}(p) \equiv D_{\L\L_0}(0,p)$.

Observe that $\tilde{c}_1^{(1)}$ would be ill-defined for $\L_R = 0$
due to an infrared divergence. 
This is the simplest indication that it is not possible to impose the
mass-independent renormalization condition 
$\partial_{m^2}|_{0} \Gamma_2^{I \L=0 \L_0}(m;\mu) = 0$, as discussed
in the previous subsection.

The one and two-loop contributions to the
four-point relevant Wilsonian vertex at scale $\L$ are  given
respectively by:
\bey\label{G4}
&&\Gamma^{\L\L_0 (1)}_4(0;0,0,0) = {-3\over 2}g^2 \int_{q}
[D_{\L\L_0}^2 - D_{\L_R\L_0}^2](q) \nonumber \\
&&\Gamma^{\L\L_0 (2)}_4(0;0,0,0) = 3g^3 \int_{pq}  D_{\L\L_0}^2(q)
[ D_{\L\L_0}(p)D_{\L\L_0}(q+p) - D_{\L_R\L_0}^2(p) ]+ \nonumber \\
&&{3 \over 4}g^3 [ \int_p (D_{\L\L_0}^2-D_{\L_R\L_0}^2)(p)]^2
+{3 \over 2}g^3 \int_{pq}  [(D_{\L\L_0}-D_{0\L_0})(p)
D_{\L\L_0}^3(q)] + g^3A({\L_R \over \L_0})
\eny
where 
$g^3A({\L_R \over \L_0})=c^{(2)}_3-{3\over 4}g^3[\int_p D_{\L_R\L_0}^2(p)]^2$ 
is the constant necessary to satisfy the 
renormalization condition (\ref{renG1}).

The other non-trivial relevant couplings we need to compute are
\bea\label{g2m}
\partial_{m^2}|_{0} \Gamma_2^{\L\L_0(1)}(m;0) ={g\over 2} \int_q 
\partial_{m^2}|_{0}[ D_{\L\L_0}(m;q) -  D_{\L_R\L_0}(m;q)]
\ena
and
\bea\label{g2}
\partial_{p^2}|_0 \Gamma^{\L\L_0 (2)}_2(0;p) = 
{-g^2 \over 6} \int_{qr} D_{\L\L_0}(q)  
D_{\L\L_0}(r) {1 \over 8} { \partial^2 \over \partial
p_{\mu} \partial p_{\mu}}|_{0}D_{\L\L_0}(p+q+r) +c_2^{(2)}
\ena
Let us consider the class of HS schemes characterized by
a propagator of the form
\bea\label{prop}
D_{\L\L_0}(m;p) = \int_{\L_0^{-2}}^{\infty} d \alpha e^{-\alpha (p^2+m^2)} \rho
(\alpha \L^2 )
\ena
where the function $\rho(x)$ satisfies $\rho(0) = 1~,~\rho'(0) =0$ and
goes sufficiently fast to zero for $x \to \infty$.
The first  condition guarantees that the hard propagator satisfies 
$D_{\L\infty}(m;p) \simeq D(m;p)$ 
for $p^2 >> \L^2$ ; the second condition (which will
not be used in the following computations) is chosen in such
a way that for $\L_0 \to \infty$ the soft propagator 
$D_S(m;p)=DK_{\L}(m;p)$ goes to
zero at least as fast as 
${\L^4 \over (p^2+m^2)^3}$ for $p^2 >> \L^2$.
For instance $\rho(x)= \theta (1-x)$ leads to
$K_{\L}(m;p)=e^{-(p^2+m^2)/\L^2}$ 
and $\rho(x)= (1+x)e^{-x}$ leads to $K_{\L}(m;p)= {\L^4 \over
(p^2+m^2+\L^2)^2}$, which
have been used in HS schemes in \cite{El} and \cite{PRR} respectively.

At one loop we get
\bea
\gamma_m^{(1)}= -\beta^{(1)}_1 = \lim_{\L_0 \to \infty} 
{g \over 16 \pi^2}\int_{{\L_R^2 \over \L_0^2} }^{\infty}
d\alpha {d \rho \over d\alpha} = {-g \over 16 \pi^2}
 \nonumber
\ena
and
\bey
\beta^{(1)} = \beta^{(1)}_3
 = \lim_{\L_0 \to \infty} 
{-6g^2 \over 16 \pi^2} \int_{{\L_R^2 \over \L_0^2} }^{\infty}
{d\alpha_1 d\alpha_2 \over (\alpha_1 +\alpha_2)^2} 
 \alpha_1 {d \rho \over d\alpha_1}  \rho (\alpha_2) =
{3g^2 \over 16\pi^2} \nonumber
\eny
where in the last step we have integrated by parts and used $\rho(0)=1$.

The last integral in (\ref{G4}) is constant in $\L$ in the limit
of infinite ultraviolet cut-off, so that eqs.(\ref{Wbeta},\ref{G4}) give
\bey\label{g4}
\beta^{(2)}_3 =&& \lim_{\L_0 \to \infty} 
\L {\partial \over \partial \L } |_{\L_R} 
3g^3 \int_{pq}  D_{\L\L_0}^2(q) [ D_{\L\L_0}(p)D_{\L\L_0}(q+p) -
D_{\L_R\L_0}^2(p) ]  \nonumber \\
=&& \lim_{\L_0 \to \infty}
{6g^3 \over (16 \pi^2)^2} \int_{{\L_R^2 \over \L_0^2} }^{\infty}
[\Pi_1^4d\alpha_i] f(\alpha_1,\alpha_2,\alpha_3,\alpha_4)
\Sigma_1^4 \alpha_i {d \over d\alpha_i} [\Pi_1^4 \rho (\alpha_j)]
  \nonumber \\
f(\alpha_1,\alpha_2,&&\alpha_3,\alpha_4) \equiv 
{1 \over [(\alpha_1+\alpha_2)(\alpha_3+\alpha_4)+\alpha_3\alpha_4]^2} -
{ 1 \over 2 (\alpha_1+\alpha_2)^2 (\alpha_3+\alpha_4)^2 } \nonumber
\eny
Integrating by parts, using $\rho(0)=1$ and the homogeneity of the
rational function $f$ in the above formula, we get
\bey
\beta^{(2)}_3 = 
{-12g^3 \over (16\pi^2)^2} \lim_{\epsilon \to 0} \int_1^{\infty} dx dy dz
\rho (\epsilon x)\rho (\epsilon y)\rho (\epsilon z)
[f(1,x,y,z)+f(x,y,1,z)] = {-6g^3 \over (16\pi^2)^2} \nonumber
\eny
where, after reordering the terms in the integrand, the limit can be
taken inside the integral, which is reduced to elementary integrals
\bey
\int_1^{\infty} {dx dy dz \over [(1+x)(y+z)+yz]^2}
+ \int_1^{\infty} dx dy dz 
{1\over (1+x)(y+z)+x]^2} - {1\over (1+x)^2(y+z)^2} ={1\over 2} \nonumber
\eny 
From (\ref{beta},\ref{Wbeta}) and (\ref{g2})
\bey
\gamma_{\phi}^{(2)} = {-\beta^{(2)}_2\over 2}=\lim_{\L_0 \to \infty} 
{-g^2 \over 4(16 \pi^2)^2} \int_{{\L_R^2 \over \L_0^2} }^{\infty}
[\Pi_1^3d\alpha_i]
\rho (\alpha_1) \alpha_2 {d\rho \over d\alpha_2} \rho (\alpha_3)
{d \over d\alpha_3} [{\alpha_3 \over
(\alpha_2+\alpha_3)\alpha_1+\alpha_2\alpha_3}]^2  \nonumber
\eny
Integrating by parts, using $\rho(0)=1$ and eq.(\ref{beta}) 
we get the standard results
\cite{ZJ}
\bea
\gamma_{\phi}^{(2)}={g^2 \over 12(16 \pi^2)^2}~~;~~
\beta^{(2)} = {-17 \over 3} { g^3 \over (16\pi^2)^2}
\ena
It is easy to see that the gradient flow property fails at three
loops;
in fact from (\ref{beta}) it follows that
\bea
\beta^{(3)} =\beta_3^{(3)}-2g\beta_2^{(3)}-{1\over 2}\alpha^{(1)}_3\beta_2^{(2)}
\ena
where
\bea
\alpha^{(1)}_3= -6 g^2 \int_p {K_{\L_R}(p)(1-K_{\L_R}(p))^2  \over (p^2)^2}
\ena
is non-vanishing for a generic cut-off function.

Let us finally discuss the case of the renormalization conditions
(\ref{renG3}), which apply only to the massive theory; the beta and
gamma functions are mass-dependent; at the first two loops $\beta$ and
$\gamma_{\phi}$ coincide, in the massless limit, with those given
above.
The function $\gamma_m$  in (\ref{BETAA}) does not admit for a generic
cut-off the massless limit; in fact
\bea
\gamma_m^{(1)} = {g \over 16\pi^2 m^2} \int_0^{\infty} {q^3 \over q^2+m^2}
\L {\partial \over \partial \L } |_{\L_R} K_{\L}(m;q)
\ena
has a singular term for $m \to 0$ \cite{HL}.

\section{Proof of the renormalization group equation in the
massless case}

In this section we begin with a short review on the renormalization
and infrared finiteness in the massless case, including the results
concerning composite operators, in the framework of the flow equation.
Some of the proofs needed are given in Appendixes A and B, as the
particular case $m=0$ of the massive theory defined in Section III.
Using the same technique we will prove the renormalization group equation in
the massless case.
The flow equation technique requires some restriction on the form of
the hard propagator; in \cite{Polchi,KKS} a smooth cut-off function with
compact support is used; another possible choice is the exponential
cut-off \cite{BT} (corresponding to the choice 
$\rho (x) =\theta (1-x)$ in eq.(\ref{prop})).

 From now on we will use the cut-off function
$K_{\Lambda}(m,p)=K\left({p^2+m^2\over \L^2}\right)$
where $K(x)$ is a smooth function, satisfying
$K(x)=1$ for $x < 1$ and $K(x)=0$ for $x>4$.
The propagator with
ultraviolet cut-off $\L_0$ and running cut-off $\L$ will be chosen of the form
\bea
D_{\L \L_0}(m,p)= D(m,p) (K_{\L_0}(m;p)-K_{\L}(m;p) )
\ena
where $D(m;p) = 1/(p^2+m^2)$ is the propagator.

Introduce $L_{\L\L_0}$ satisfying
\bea\label{L}
e^{-{1\over \hbar}L_{\L\L_0}[\Phi]} = 
e^{\hbar \Delta_{\L\L_0}}e^{-{1\over \hbar}L_{\L_0}[\Phi]}
\ena
with
\bea\label{Lbare}
L_{\L_0}[\Phi] = S^I(\Phi)
\ena
where $S^I$ is the interacting part of the bare action 
given in eqs. (\ref{barem},\ref{MMM}) with $m=0$, and
\bea
\Delta_{\L\L_0} = {1 \over 2} {\delta \over \delta \Phi} D_{\L\L_0}
{\delta \over \delta \Phi}
\ena
The Wilsonian effective action  (\ref{Zj}) is given by \cite{KKS}
\bea\label{ZP}
Z_{\L\L_0}[J] = e^{{1\over \hbar}W_{\L\L_0}[J]} =
e^{{1\over \hbar}\omega_{\L\L_0}[D_{\L\L_0}J]}
\ena
where
\bea\label{om}
\omega_{\L\L_0}[\Phi] \equiv {1 \over 2} \Phi D_{\L\L_0}^{-1}\Phi - 
L_{\L\L_0}[\Phi]
\ena

$L_{\L\L_0}[\Phi]$ is the functional generator of the truncated connected
Green functions (apart from the tree-level two-point function) of the
Wilsonian theory with propagator $D_{\L\L_0}$ and bare vertices $S^I$.
Formulae (\ref{L}) and (\ref{ZP}) are understood as relations between
formal power series in $\hbar$ in which only the terms depending on
the fields are well-behaved. The same considerations apply to all
following functionals; therefore all the formulae hold up to terms
constant in the fields.

$L_{\L\L_0}$ satisfies the flow equation
\bea\label{flux}
\partial_\Lambda L_{\Lambda\L_0} =
\hbar \left(\partial_\Lambda \Delta_{\Lambda
\Lambda_0}\right)L_{\Lambda\L_0} -{ 1 \over 2}
L'_{\Lambda\L_0}\left(\partial_{\L}D_{\Lambda\Lambda_0}\right)
L'_{\Lambda\L_0}
\ena
where $L'_{\Lambda\L_0} \equiv {\delta L_{\Lambda\L_0}\over\delta \Phi}$.
The flow equation can be solved in perturbation theory assigning the
relevant couplings at $\L = \L_R$.
Define the relevant couplings at scale $\L > 0$:
\bea\label{rel}
\rho_1^{\L\L_0} = L^{\L\L_0}_2(0) ~~;~~\rho_2^{\L\L_0} = 
\partial_{p^2}L^{\L\L_0}_2|_{p=0}~~;~~
\rho_3^{\L\L_0} = L^{\L\L_0}_4|_{p_i=0}
\ena
All the other couplings are irrelevant and are fixed to be zero at 
$\L= \L_0$ by eq.(\ref{Lbare}) .  Using the flow equation one can compute
the connected Green functions at scale $\L$ using these boundary
conditions;
indeed  conditions (\ref{rel}) and  (\ref{Lbare}) are in
a biunivocal relation in perturbation theory.
 This observation is the starting point for proving
renormalizability with the method of the flow equation \cite{Polchi}.
While the relevant couplings $\rho_2^{\L\L_0}$ and $\rho_3^{\L\L_0}$
are defined only for $\L > 0$, $\rho_1^{\L\L_0}$ can be extended to 
$\L = 0$, and it must vanish there to avoid infrared divergences in
the effective Green functions $L_n^{\L\L_0}(p_1,...,p_{n-1})$ for
non-exceptional momenta in the limit $\L \to 0$. This limit will be
studied in detail in the next section and in  Appendix B, as the
particular case in which the mass $m$ is taken to be zero.

Choose the renormalization conditions
\bea\label{rc}
\rho_1^{\L=0\L_0(l)} = 0~~;~~\rho_2^{\L_R\L_0(l)} = 0 ~~;~~
\rho_3^{\L_R\L_0(l)} = g~\delta^{l,0}
\ena
which are the same as those chosen in the first section in 
eqs.(\ref{renG0}-\ref{renG2}) (this is true
only for compact-support cut-off; for an analytic cut-off there is an
extra term in the renormalization condition on $\rho_2$ in (\ref{rc}),
to get the renormalization conditions (\ref{renG1}) on the two-point vertex).

In order to prove renormalizability with the flow equation technique,
an appropriate norm must be defined.

Given a function $f(m;p_1,...,p_{n-1})$, and  $p_n \equiv -\sum_1^{n-1}p_i$,
we define the norm
\bea\label{norM}
||f||_{\L}= Sup_M |f|
\ena
where $M = \{p,m: |p_i| \leq Max (\eta, 2\L ), i=1,..,n,~m \leq \eta \}$
and $\eta$ is a fixed positive quantity. In this section the mass $m$
is zero.

In Appendix A we will prove that
\bea\label{rtz}
||\partial_{\L}^w \partial_p^z L^{\L\L_0}_n ||_{\L}
\leq \L^{4-n-w-z} P(ln {\L \over \L_R})
\ena
where $P$ is a polynomial;
using this bound it is easy to prove renormalizability.

Similarly one can study the functionals representing the insertion of
an operator.
In our framework the flowing functional generator of the connected and
amputated Green functions with one insertion of an operator $\cal O$
is \cite{KKop}
\bea\label{fluxO2}
{\cal O}_{\L\L_0}[\Phi] =  e^{{1\over \hbar}L_{\L\L_0}[\Phi]} 
e^{\hbar \Delta_{\L\L_0} ({\delta \over \delta \Phi})} {\cal O}^{\L_0}[\Phi]
 e^{-{1\over \hbar}L_
{\L_0}[\Phi]}
\ena
which satisfies
\bea\label{fluxO}
\partial_\Lambda {\cal O}_{\Lambda\L_0} =
\hbar \left(\partial_\Lambda \Delta_{\L\L_0}\right){\cal O}_{\Lambda\L_0} -
L'_{\Lambda\L_0}\left(\partial_{\L}D_{\Lambda\Lambda_0}\right)
{\cal O}'_{\Lambda\L_0}
\ena
${\cal O}^{\L\L_0}[\Phi]$ interpolates between the bare operator 
${\cal O}^{\L_0}[\Phi]$, boundary condition of eq.(\ref{fluxO}) and
the physical functional generator of the insertion ${\cal
O}^{0\L_0}[\Phi]$.
An operator of dimension $D$ is associated to boundary conditions 
on ${\cal O}^{\L_0}_n(p_1,...,p_{n-1})$ such that
\bea\label{bou}
|| \partial_{p}^z {\cal O}_n^{\L_0(l)}||_{\L_0} \leq \L_0^{D-n-z}
P(ln {\L_0 \over \L_R })
\ena
for $D-n-z < 0$ (see \cite{KKop}), $P$ being a polynomial,
and satisfies renormalization
conditions at $\L_R$ on its relevant part, defined at vanishing
momentum and at scale $\L = \L_R$, up to dimension $D$.

In this paper we are interested only in the case of integrated scalar
composite operators of dimension four.
The simplest choice for the boundary condition at  $\L = \L_0$ would be 
${\cal O}^{\L_0}_n = 0$ for $D-n-z < 0$, but for our purposes it is
useful to keep the more general condition  (\ref{bou}).
The renormalization conditions for a $D=4$ operator have the form
\bey\label{renO}
{\cal O}^{\L_R\L_0 (l)}_2(0) = a^{(l)}_1~~;~~
\partial_{p^2}|_{p=0} {\cal O}^{\L_R\L_0 (l)}_2(p) =a_2^{(l)}~~;~~ 
{\cal O}^{\L_R\L_0 (l)}_4|_0 = a_3^{(l)}
\eny
In order to guarantee the existence of the limit $\L \to 0$, the 
coefficients $a^{(l)}_1$ must be tuned in such a way that 
\bea\label{renO2}
{\cal O}^{0\L_0 (l)}_2(0)=0
\ena

Let us now discuss the renormalization group equation.
Making the theory flow from $\L_R$ to $\L_R'$ the new relevant
couplings are functions of $\L_R$, $\L_0$ and $g$ which are determined
by the flow equation 
\bey\label{rc2}
\rho_2^{\L_R'\L_0} =\partial_{p^2}|_{p=0} L^{\L_R'\L_0}_2(\L_R,0,g;p)
~~;~~\rho_3^{\L_R'\L_0} =  L^{\L_R'\L_0}_4(\L_R,0,g;0,0,0)
\eny
while the renormalization condition on $\rho_1$ remains the same as
in (\ref{rc}).
Interpreting $\L_R'$ as the new renormalization point and 
$\rho_2^{\L_R'\L_0}$ and $\rho_3^{\L_R'\L_0}$ as the new relevant
couplings, the theory is unchanged, since these two conditions are on
the same renormalization group trajectory:
\bea\label{ER}
L^{\L\L_0}[\L_R',\rho_2^{\L_R'\L_0},\rho_3^{\L_R'\L_0};\Phi] =
L^{\L\L_0}[\L_R,0,g;\Phi]
\ena
This is an exact renormalization group equation, holding for any $\L$,
$\L_R'$ and $\L_0$.
Differentiating this equation with respect to $\L_R'$ and setting
$\L_R' = \L_R$ we get the following exact renormalization group
equation \cite{HL}:
\bea\label{ERG}
\L_R {\partial \over \partial \L_R } L^{\L\L_0}[\L_R,0,g;\Phi]
+ \beta_2 {\partial \over \partial \rho_2 }|_{\rho_2=0}
L^{\L\L_0}[\L_R,\rho_2,g;\Phi] + \beta_3{\partial \over \partial g } 
L^{\L\L_0}[\L_R,0,g;\Phi] = 0
\ena
which holds for any $\L$ and $\L_0$, with
\bea\label{b}
\beta_2^{(l)}(g,\L_R,\L_0) =  
\L {\partial \over \partial \L }|_{\L=\L_R} \rho_2^{\L\L_0(l)}~~;~~
\beta_3^{(l)}(g,\L_R,\L_0) =  
\L {\partial \over \partial \L }|_{\L=\L_R} \rho_3^{\L\L_0(l)}
\ena
This discussion on the exact renormalization group equation 
is rather formal; the equation (\ref{ERG}) must be justified
in perturbation theory; this is done in Appendix C.

Let us write equation (\ref{ERG}) in the form
\bea\label{erg}
( \L_R {\partial \over \partial \L_R } +\beta_3{\partial \over \partial g } )
L^{\L\L_0}[\L_R,0,g;\Phi] + \beta_2 {\cal O}_k^{\L\L_0}[\Phi] = 0
\ena
where
\bea\label{ok}
{\cal O}_k^{\L\L_0}[\Phi] \equiv
{\partial \over \partial \rho_2 }|_{\rho_2=0}
L^{\L\L_0}[\L_R,\rho_2,g;\Phi]
\ena

Since $\L_R$ is the mass scale parameter introduced in the renormalization
procedure, eq.(\ref{erg}) should correspond to a generalization  of
the Gell-Mann and Low renormalization group equation. To prove this
correspondence starting from (\ref{erg}) we have to obtain an equation in which
the last term ${\cal O}_k^{\L\L_0}$ is replaced by the `number operator'
$\Phi{\delta \over \delta \Phi}\omega_{\L\L_0}[\Phi]$;  besides that, we
expect that the usual renormalization group equation is recovered only in
the limits $\L_0 \to \infty$ and $\L \to 0$.
To this aim we have to prove some results concerning the Quantum Action
Principle \cite{Qap} derived in the context of the flow equation
technique in
\cite{Becchi2}. We want to prove that the functional obtained by
differentiation of the functional $\omega_{\L\L_0}[\Phi]$
with respect to its parameters can be obtained as insertions of
suitable local operators of dimension four.

Equation (\ref{erg}) can now be understood  as a relation concerning
three generators of insertions; indeed the functionals
\bea\label{o}
{\cal O}_{\L_R}^{\L\L_0}[\Phi] \equiv
\L_R {\partial \over \partial \L_R }L^{\L\L_0}[\L_R,0,g;\Phi]~~;~~
{\cal O}_g^{\L\L_0} = {\partial \over \partial g }L^{\L\L_0}[\L_R,0,g;\Phi]
\ena
and ${\cal O}_k^{\L\L_0}$ defined in  (\ref{ok}) satisfy  (\ref{fluxO}) and
 (\ref{bou}) for $D=4$, indeed only the first two  functionals are in a
direct relation with the generator $\omega^{\L\L_0}[\L_R;0,g;\Phi]$,
being ${\cal O}_{\L_R}^{\L\L_0}=-\L_R \partial_{\L_R} \omega^{\L\L_0}$ and
${\cal O}_g^{\L\L_0}=-\partial_g \omega^{\L\L_0}$, and in this sense they
explicitely agree with the Quantum Action Principle.
In order to express ${\cal O}_k^{\L\L_0}$ in a similar way we start by considering
the `number operator' $\Phi {\delta \over \delta \Phi}$ acting on the
generator of the connected and amputated Green functions
\bea\label{N1}
{\cal O}_N^{0\L_0}[\Phi] = e^{-{1\over \hbar}\omega^{0\L_0}[\Phi]}
\hbar \Phi K_{\L_0} {\delta \over \delta \Phi}
e^{{1\over \hbar}\omega^{0\L_0}[\Phi]}
\ena
where, for technical reasons, we introduced the factor $ K_{\L_0}$ in
the number operator ($K_{\L_0} \to 1$ for $\L_0 \to \infty$).
The functional ${\cal O}_N^{0\L_0}$ can be
computed as the $\L \to 0$ limit of the following functional
\bey\label{N}
{\cal O}^{\L\L_0}_N =&& e^{{1\over \hbar}L^{\L\L_0}[\Phi]} 
e^{-\hbar\Delta_{0\L}}
[ \Phi D^{-1}\Phi +\hbar \Phi K^{\L_0}{\delta \over \delta \Phi} ]
e^{-{1\over \hbar}L^{0\L_0}[\Phi]} \nonumber \\
=&& e^{-{1\over \hbar}\omega^{\L\L_0}}
[\hbar \Phi K^{\L_0}{\delta \over \delta \Phi}
-\hbar^2{\delta \over \delta \Phi}D_{\L\L_0}K_{\L}{\delta \over \delta \Phi}]
e^{{1\over \hbar}\omega^{\L\L_0}}
\eny
which solves, because of its definition, the flow equation of an
operator insertion  (\ref{fluxO}). Expanding the above expression we
get
\bea\label{N2}
{\cal O}^{\L\L_0}_N = \Phi D^{-1}\Phi - L_{\L\L_0}' D_{\L\L_0}
K_{\L}L_{\L\L_0}' + \hbar tr D_{\L\L_0}K_{\L}L_{\L\L_0}''
-\Phi (K_{\L_0}-2K_{\L}) L_{\L\L_0}' 
\ena
Using this equation for $\L = \L_0$ and the bounds (\ref{rtz}) satisfied by
$L_{\L\L_0}$ one checks the validity of eq.(\ref{bou}) for $D=4$. 

From (\ref{N2}) and  (\ref{rc}) one gets
\bea\label{rcN}
{\cal O}^{\L=0\L_0}_{N~2}(0)=0~~;~
\partial_{p^2} {\cal O}_{N~2}^{\L_R\L_0}|_{p=0}=2+\alpha_2~~;~~
{\cal O}^{\L_R\L_0}_{N~4}|_{p_i=0}=4g+\alpha_3
\ena
where
\bey\label{aaa}
&&\alpha_2 =\hbar \partial_{p^2}|_0 \int_q D_{\L_R\L_0}(q) K_{\L_R}(q)
L^{\L_R\L_0}_4(q,-q,p) \nonumber \\
&&\alpha_3 =\hbar \int_q D_{\L_R\L_0}(q) K_{\L_R}(q)
L^{\L_R\L_0}_6(q,-q,0,0,0)
\eny
are the same quantities defined in (\ref{a2}) (as above,
this correspondence is exact only for a compact-support cut-off function).
We can conclude that the functional   (\ref{N}) can be computed as the
insertion of a `local operator' of dimension $D=4$ whose renormalization
conditions are given in  (\ref{rcN}).

The same considerations apply in a simpler way to the other two
insertions appearing in eq.  (\ref{erg}).
For ${\cal O}_k^{\L\L_0}$ the renormalization conditions are
\bea
{\cal O}^{\L=0\L_0}_{k~2}(0)=0~~;~~
\partial_{p^2}{\cal O}^{\L_R\L_0}_{k~2}|_{p=0}=1~~;~~
{\cal O}^{\L_R\L_0}_{k~4}|_{p_i=0}=0
\ena
so that ${\cal O}_k^{\L\L_0}$ is the operator insertion which at
tree-level is equal to $\Phi D^{-1}\Phi$; for the other operator,
\bea
{\cal O}^{\L=0\L_0}_{g~2}(0)=0~~;~
\partial_{p^2} {\cal O}_{g~2}^{\L_R\L_0}|_{p=0}=0~~;~~
{\cal O}^{\L_R\L_0}_{g~4}|_{p_i=0}=1
\ena
so that ${\cal O}_g^{\L\L_0}$ is the operator insertion which at
tree-level is equal to $\Phi^4$.

A generic operator ${\cal O}^{\L\L_0}$ of dimension four,
satisfying the renormalization conditions
\bea
{\cal O}^{\L=0\L_0}_{2}(0)=0~~;~
\partial_{p^2} {\cal O}_{2}^{\L_R\L_0}|_{p=0}=a~~;~~
{\cal O}^{\L_R\L_0}_{4}|_{p_i=0}=b
\ena
can be decomposed in terms of ${\cal O}_k^{\L\L_0}$ and ${\cal
O}_g^{\L\L_0}$; in fact
\bea
{\cal E}^{\L\L_0} \equiv {\cal O}^{\L\L_0}-a{\cal O}_k^{\L\L_0}-b{\cal O}_g^{\L\L_0}
\ena
is an operator insertion satisfying eq.(\ref{fluxO}), the bounds (\ref{bou})
and it has vanishing renormalization conditions
\bea
{\cal E}^{\L=0\L_0}_{2}(0)=0~~;~
\partial_{p^2} {\cal E}_{2}^{\L_R\L_0}|_{p=0}=0~~;~~
{\cal E}^{\L_R\L_0}_{4}|_{p_i=0}=0
\ena
so that it is evanescent (see \cite{KKop} and Appendix A).

Therefore  (\ref{rcN}) implies
\bea\label{N4}
{\cal O}^{\L\L_0}_{N} = (2+\alpha_2){\cal O}^{\L\L_0}_k
+(4g + \alpha_3) {\cal O}^{\L\L_0}_g + {\cal E}^{\L\L_0}
\ena
where ${\cal E}^{\L\L_0}$ is evanescent.

Eq.(\ref{N4}) is the relation expressing ${\cal O}_k^{\L\L_0}$ in
terms of $L^{\L\L_0}$ (or $\omega^{\L\L_0}$) which we were looking for.
Introducing in  (\ref{erg}) the expression for ${\cal O}_k^{\L\L_0}$
in eq.(\ref{N4}) we get
\bea\label{rgez}
[ \L_R {\partial \over \partial \L_R } +
\beta {\partial \over \partial g }]\omega^{\L\L_0}
 +\gamma_{\phi}[{\cal O}_N^{\L\L_0} -{\cal
E}^{\L\L_0}] =0
\ena
where
\bey\label{BETA}
\beta = \beta_3 + \gamma_{\phi} (4g+\alpha_3)~~;~~~~
\gamma_{\phi} = {-\beta_2 \over 2+\alpha_2}
\eny
with $\beta_2, \beta_3$ defined in (\ref{b}) and $\alpha_2, \alpha_3$
defined in (\ref{aaa}). 
Using (\ref{N}) we obtain
\bea\label{rgom}
[ \L_R {\partial \over \partial \L_R } +
\beta {\partial \over \partial g } +
 \gamma_{\phi} (\Phi K^{\L_0} {\delta \over \delta \Phi}
-\hbar {\delta \over \delta \Phi} D_{\L\L_0} K^{\L} {\delta \over \delta
\Phi}
- {\cal E}^{\L\L_0}[\Phi])]
e^{{1\over \hbar}\omega^{\L\L_0}[\Phi]} = 0
\ena

Notice that in the limit $\L_0 \to \infty$ $\beta$ and $\gamma_{\phi}$ have 
the same expressions as in eq.(\ref{beta}) (in the limit of cut-off
function with compact support); using (\ref{ZP}) 
in (\ref{rgom}) we obtain the effective renormalization group equation
(\ref{rgef}) (without the mass term), as we claimed.
In particular, in the limit $\L \to 0$ we get the
 renormalization group equation
\bea
( \L_R {\partial \over \partial \L_R } +
\beta (g) {\partial \over \partial g } +
\gamma_{\phi} (g) J {\delta \over \delta J}) Z[J] = 0
\ena
The existence of this limit will be proven in the next section and in
Appendix B.

\section{Mass-independent renormalization and the renormalization group}

The discussion of the (effective) renormalization group equation of
the previous section could be straightforwardly extended to the
massive case, however in a generic scheme one would get a dependence
on $\L_R \over m$ of the $\beta$ and $\gamma$ functions.
This is an unwanted complication from the point of view of the
utilization of the  renormalization group equation in 
studying the asymptotic behavior of the Green functions.
The mass-independent renormalization scheme introduced in \cite{wein2}
overcomes this problem but it has some difficulties in the theories
with scalars due to infrared divergences. In later treatments \cite{BZ}
the problem is in some way solved by defining a generalized functional
of the massless theory with a new source $\int k(x) \phi^2(x) dx$ ,
which is supposed to be summed exactly to all orders in $k(x)$, making
at the end the replacement $k(x) \to m^2$. In this way the infrared
singularities which are present for a finite number of $\int \phi^2(x) dx$ 
insertions are avoided. In \cite{GS} a mass-independent approach
similar to the HS approach originally proposed in \cite{LM} is
developed, but it applies specifically to massive theories, while the
HS schemes apply equally well to massless theories.

In this section we consider a mass-independent HS scheme, outlined in
Section I,
in which the above-mentioned infrared problems are overcome in a more 
natural way: instead of defining mass-insertions, it will be
sufficient to make derivatives with respect to the mass; the
mass parameter will be considered more like a `momentum'  than as a
coupling constant. The mass-independent renormalization group equation
is then obtained in a rather straightforward way. Technical details of
the proof are postponed to the appendixes.

The bare coefficients
 $\hat{c}_1, {\tilde c}_1, c_2, c_3$ in the bare action 
(\ref{barem},\ref{MMM}) do not depend on $m^2$.
Perturbatively these parameters are
in an invertible relation with the following mass-independent
renormalization conditions in a formal loop expansion:
\bey\label{renLm}
&&L^{ \L_R\L_0(l)}_2(0;0) = \rho^{(l)}_1~~;~~~~
\partial_{m^2}|_{m=0} L^{ \L_R\L_0(l)}_2(m^2;0) ={\tilde \rho}^{(l)}_1
 \nonumber \\
&&\partial_{p^2}|_{p=0} L^{ \L_R\L_0(l)}_2(0;p) =\rho_2^{(l)}~~;~~ 
L^{ \L_R\L_0(l)}_4 (0;0,0,0) = \rho_3^{(l)}
\eny
with $\rho_1^{(0)} ={\tilde \rho}_1^{(0)} =\rho_2^{(0)} = 0$ (so that
 $L^{(0)\L\L_0}_2= 0$).  

We have to prove that eqs.(\ref{renLm}) lead to ultraviolet
renormalization. The proof is a modification of the one made in
\cite{KKS}, in which one considers derivatives with respect to the mass
besides those with respect to the momenta. Indeed
the functions $L^{\L\L_0}_n(m;p_1,...,p_{n-1})$ for $\L > 0$ and
$\L_0 < \infty$ are, due to (\ref{barem},\ref{MMM},\ref{renLm}), 
$C^{\infty}$ functions of
$n-1$ independent momenta and also of the mass $m$. One can prove by
induction in the loop index $l$ and in the number of external legs $n$
the following bounds ( $\L \geq \L_R$ and $z$ is the total order of
the derivatives with respect to the momenta):
\bea\label{rt}
||\partial_{\L}^w \partial_{m^2}^r \partial_p^z L^{\L\L_0}_n ||_{\L}
\leq \L^{4-n-w-2r-z} P(ln {\L \over \L_R})
\ena
where $w=0,1$ and $P$ is a polynomial with positive coefficients which
are independent from $\L,\L_0$ and $m^2$; in general $P$ will denote a
different polynomial whenever it appears.

Eq.(\ref{rt}) is essentially the renormalization theorem; using it one
can show that
\bea\label{BOU}
||\partial_{\L_0} \partial_{m^2}^r \partial_p^z L^{\L\L_0 (l)}_n   ||_{\L}
\leq {\L^{5-n-z-2r} \over \L_0^2}P(ln {\L_0 \over \L_R})
\ena
and from this the proof of the existence of $\lim_{\L_0 \to \infty}
\partial_{m^2}^r \partial_p^z L^{\L\L_0 (l)}_n$ follows easily
(uniformly on the compact sets of momenta and mass). The details of
the proof are given in appendix A.
The only difference with the quoted standard proof is that for
the relevant interacting Green function we use Taylor reconstruction with
respect to the mass too.

The theory at $\L=0$ is obtained collecting graphs
with $L^{\L_R\L_0}_n$ vertices and soft propagator $D_{0\L_R}(m;p)$,
which has compact support. From the existence of 
$\lim_{\L_0 \to \infty}L^{\L_R\L_0 (l)}_n$
it follows that 
$\lim_{\L_0 \to \infty}L^{0\L_0 (l)}_n$ exists too, and
 is a $C^{\infty}$ function of the momenta for $m>0$.
In order that the parameters of the Gell-Mann and Low equation for the
massive case be mass-independent, they must be the same as in the
massless theory, so that we want to show that
$\lim_{m \to 0}L^{0\L_0 (l)}_n(m;p_1,...,p_{n-1})
=L^{0\L_0 (l)}_n(0;p_1,...,p_{n-1})$ are the Green functions
considered in the previous section.
On the other hand in Section II we did not prove that
$\lim_{\L \to 0}L^{\L\infty (l)}_n(0;p_1,...,p_{n-1})$
exists for non-exceptional momenta.
This fact could be proven directly, however we find it more economical
to solve together these two problems by
proving the existence of the global limit $(\L,m,\L_0) \to
(0,0,\infty)$
of the Green functions for non-exceptional momenta. 

In the course of the proof, using the flow
equation, it will be necessary to establish bounds on Green functions with
some sets of $k$ exceptional momenta, with $0 \leq k \leq n-1$.

Let $\eta_1$ and $\eta_2$ be fixed positive quantities. 
We want to consider the sets
of $n-1$ independent momenta $p_i$, $k$ of which are `small' (i.e. 
$|p_i| < 2\L$), while the remaining $n-1-k$ momenta, and all their
partial sums, are larger than $\eta_1$. We will not need to use the
more general sets of momenta, in which a subset of momenta is large,
but partial sums of large momenta are not necessarily large.
In this sense our treatment is similar to the infrared treatment of
the massless case in \cite{BDM}. More general sets of exceptional
momenta have been considered in \cite{KK}.

Let us define norms $||~~~||_{\L}^k$, for $0 \leq k \leq n-1$.
Define for $k$ with $0 \leq k < n-1$ the following norm for a 
smooth function $f$
of $n-1$ independent momenta $\{p_1,...p_{n-1} \}$
\bea
||\partial_p^{z,z'}f||_{\L}^k \equiv Max Sup_{X(k,\L)} 
|{\partial^{z_1} \over \partial p_{i_1}} ...
 {\partial^{z_k} \over \partial p_{i_k}}
{\partial^{z_{k+1}} \over \partial p_{i_{k+1}}}..
{\partial^{z_{n-1}} \over \partial p_{i_{n-1}}} f|
\ena
where $Max$ is taken over $\sum_1^k z_i =z$ and 
$\sum_{k+1}^{n-1} z_i =z'$ and over 
the permutations of the first $n-1$
integers, whose generic element is called $i_1,...,i_{n-1}$, and the set
$X(k,\L)$ is so defined:
\bey\label{set}
&&X(k,\L)\equiv \{p_1....p_{n-1} : |p_{i_r}| < 2 \L, ~r=1,...,k;
\eta_1 < |p_{i_s}|< \eta_2, ~ s \geq k+1; \nonumber \\
&&|p_{i_{s_1}}+ p_{i_{s_2}} | > \eta_1, ~s_1,s_2 \geq k+1;~ 
|p_{i_{s_1}}+ p_{i_{s_2}}+p_{i_{s_3}}|> \eta_1, ~s_1,s_2,s_3 \geq
k+1;... \nonumber \\
&&|p_{i_{s_1}}+...+ p_{i_{s_{n-1-k}}}|> \eta_1, ~s_1,...,s_{n-1-k} \geq k+1 \}
\eny

For $k=n-1$, define the norm
\bea
||\partial_p^z f||_{\L}^{n-1} \equiv Max Sup |\partial_p^z f|
\ena
where $Max$ is the same as above and $Sup$ is taken over 
$\{p_1....p_{n-1} : |p_i| < 2 \L ;| \sum_i^{n-1}p_i| < 2 \L \}$.
For $k=0$ the index $\L$ in $||\partial_p^{z'} f||_{\L}^{0}$ is
meaningless and it can be omitted.
Using these norms, in the appendix we will prove the following theorem:
it is possible loop by loop to choose  coefficients 
$\rho^{(l)}_1 = \L_R^2 f^{(l)}(g,{\L_R \over \L_0})$ in
such a way that 
\bey\label{ren2m}
\lim_{\L \to 0} L^{\L\L_0 (l)}_2(0;0) =0
\eny
Furthermore, for arbitrary values of $\L$ and $m$ (with $\L < \L_R$) 
the following bounds hold:
consider $w=0,1$ and let $z$ be the overall order of the
derivatives with respect to the momenta; then
\bey\label{bs}
&&(\L+ {m \over 2})^{4-n-z-2r-w}{\cal{P}} ~~~~~~~~~~ k=n-1 \nonumber \\~~~~~~~~~~~
||\partial_{\L}^w \partial_{m^2}^r \partial_p^{z,z'} L^{\L\L_0 (l)}_n
||_{\L}^k \leq ~~\Big\{ ~~&& 
(\L+ {m \over 2})^{-2[{k-1 \over 2}]-z-2r-w}{\cal{P}}~~~~~~~0<k<n-1
~~~~~~~~~~~~~ \nonumber \\
&& const.  ~~~~~~~~~~~~~~~~~~~~~~~~~~k=0, ~w=r=0 \\
&&(\L+ {m \over 2})^{2-w-2r}{\cal{P}} ~~~~~~~~~~~~ k=0, ~w+r>0  \nonumber
\eny
where ${\cal{P}} \equiv P(ln {\L_R+ {m\over 2} \over \L + {m\over
2}})$ is a polynomial whose coefficients do not depend on $\L$ and $m$
and where $[x]$ represents the integer part of $x$.

 From (\ref{BOU}) and (\ref{bs})
ultraviolet convergence and  infrared finiteness for non-exceptional
momenta immediately follow; indeed
\bey\label{tz}
&& | \partial_p^{z'} L^{\L_1 \L_0}_n(m_1;p_i) - \partial_p^{z'}
L^{\L_2\L_0'}_n(m_2;p_i) | \leq
\int_{\L_1}^{\L_2} d\L' | \partial_{\L'}  \partial_p^{z'}
L^{\L'\L_0'}_n(m_2;p_i) | +  \nonumber \\ 
&& \int_{m_1^2}^{m_2^2} dm'^2| {\partial \over \partial m'^2} \partial_p^{z'}
 L^{\L_1\L_0'}_n(m';p_i) |
+ \int_{\L_{0}}^{\L_{0}'} d\L | {\partial \over \partial \L}
\partial_p^{z'} L^{\L_1\L}_n(m_1;p_i) | \to 0 
\eny
for $(\L_{1,2},m_{1,2},\L_0,\L_0') \to(0,0,\infty ,\infty)$.

One can discuss the renormalization of the
generating functional of a composite operator which satisfies
(\ref{bou}) in an analogous way. 
In this paper we are actually interested only in integrated composite
operators. In a mass-independent scheme one starts with a bare
operator ${\cal O}^{\L_0}$ which, like $L^{\L_0}$, has a local term
of dimension $D$ which depends polynomially on $m^2$.
In general there is also a
remainder of irrelevant terms, namely terms of  ${\cal O}^{\L_0}$
which are generally non-local  and are $C^{\infty}$
functions of $m^2$ and of the momenta, but which in general do not
depend in a polynomial way on $m^2$ and such that the following
inequality is satisfied
\bea\label{bob}
|| \partial_{m^2}^r \partial_p^z {\cal O}^{\L_0}_n ||_{\L_0} \leq
\L_0^{D-n-2r-z} P (ln {\L_0 \over \L_R})
\ena
for $D-n-2r-z <0$.
  $ {\cal O}^{\L\L_0}_n(m;p_1,...,p_{n-1})$ will be
$C^{\infty}$ functions of the momenta and of $m^2$ for $\L >0$. The
relevant part of  $ {\cal O}^{\L_0}$ is in a biunivocal relation with
the following renormalizaton conditions, which we write down only in
the case $D=4$, the generalization to operators of arbitrary dimensions
being obvious:

\bey\label{renOm}
&&{\cal O}^{\L_R\L_0 (l)}_2(0;0) = a^{(l)}_1~~;~~~~
\partial_{m^2}|_{m=0} {\cal O}^{\L_R\L_0 (l)}_2(m;0) ={\tilde a}^{(l)}_1 \nonumber \\
&&\partial_{p^2}|_{p=0} {\cal O}^{\L_R\L_0 (l)}_2(0;p) =a_2^{(l)}~~;~~ 
{\cal O}^{\L_R\L_0 (l)}_4 (0;0,0,0) = a_3^{(l)}
\eny
From (\ref{renOm}) the ultraviolet part of the proof follows along the
lines discussed for $L^{\L\L_0}$. In particular for an operator of
dimension $D$, for $\L \geq \L_R$,
\bea\label{bom}
||\partial_{m^2}^r \partial_p^z {\cal O}^{\L\L_0 (l)}_n ||_{\L} 
\leq  \L^{D-n-z-2r}P(ln {\L \over \L_R}) 
\ena
If $D >2$, by choosing $a_1^{(l)}$ in  (\ref{renOm}) loop by loop in
such a way that
\bea\label{O22}
\lim_{\L \to 0}  {\cal O}^{\L \L_0 (l)}_2(0;0) = 0
\ena
one can prove suitable infrared bounds which in the case $D=4$
coincides with (\ref{bs}) , and then 
the existence of $ {\cal O}_n^{\L \L_0 (l)}(m;p_1,...,p_{n-1})$ 
with non-exceptional momenta in the limit $(m,\L) \to (0,0)$ follows.

To discuss the renormalization group equation we have now to
generalize the procedure followed in (\ref{ER}). Consider standard
renormalization conditions in which the parameters in (\ref{renLm})
assume the values $\tilde{\rho}_1=\rho_2=0$ and 
$\rho_3^{(l)}=g\delta^{l,0}$, while $\rho_1$ is fixed by  (\ref{ren2m}); then 
\bea\label{ERm}
L^{\L\L_0}[\L_R';m;\tilde{\rho}_1^{\L_R'\L_0} \rho_2^{\L_R'\L_0},
\rho_3^{\L_R'\L_0};\Phi] = L^{\L\L_0}[\L_R;m;0,0,g;\Phi]
\ena
Applying $ \L_R' {\partial \over \partial \L_R' }$ in $\L_R' = \L_R$
to the left hand side of this equation we get

\bey\label{ERGm}
&&\L_R {\partial \over \partial \L_R } L^{\L\L_0}[\L_R;m,0,0,g;\Phi]
+ \beta_1 {\partial \over \partial \tilde{\rho}_1 }|_{\tilde{\rho}_1 =0}
L^{\L\L_0}[\L_R;m, \tilde{\rho}_1 ,0,g;\Phi]\nonumber \\
&&+ \beta_2 {\partial \over \partial \rho_2 }|_{\rho_2=0} 
L^{\L\L_0}[\L_R;m,0,\rho_2,g;\Phi] + 
 \beta_3{\partial \over \partial g } L^{\L\L_0}[\L_R;m;0,0,g;\Phi] = 0
\eny
where $\beta_2$ and $\beta_3$ are the same as in the massless case
(\ref{b}) and
\bea\label{bb}
\beta_1(g,\L_R,\L_0) =  
\L {\partial \over \partial \L }|_{\L=\L_R} 
 {\partial \over \partial m^2 }|_{0} L^{\L\L_0}_2[\L_R;m;0,0,g]
\ena
The exact renormalization group equation (\ref{ERm}) can be written in
the form

\bea\label{ergm}
( \L_R {\partial \over \partial \L_R } +\beta_3{\partial \over \partial g } )
L^{\L\L_0}[\L_R;m;0,0,g;\Phi] 
+ \beta_1 {\cal O}_{m^2}^{\L\L_0} + \beta_2 {\cal O}_k^{\L\L_0}= 0
\ena
where $ {\cal O}_k$ is defined as in (\ref{ok}) but with mass, while
\bea\label{om}
 {\cal O}_{m^2}^{\L\L_0} \equiv
{\partial \over \partial \tilde{\rho}_1 }|_0
L^{\L\L_0}[\L_R;m, \tilde{\rho}_1 ,0,g;\Phi]
\ena
satisfies, as well as ${\cal O}_{\L_R}$, ${\cal O}_g$ and ${\cal O}_k$
, the flow equation for operator insertions (\ref{fluxO}).
These operators, which are defined in term of $L^{\L\L_0}$, satisfy 
eq.(\ref{bob}) and the following mass-independent renormalization conditions: 
\bea\label{rcokm}
{\cal O}^{\L=0\L_0}_{k~2}(0;0)=0~~;~~
\partial_{m^2}{\cal O}^{\L_R\L_0}_{k~2}|_0=0~~;~~
\partial_{p^2}{\cal O}^{\L_R\L_0}_{k~2}|_0=1~~;~~
{\cal O}^{\L_R\L_0}_{k~4}|_0=0
\ena

\bea\label{rcogm}
{\cal O}^{\L=0\L_0}_{g~2}(0;0)=0~~;~~
\partial_{m^2}{\cal O}^{\L_R\L_0}_{g~2}|_0=0~~;~~
\partial_{p^2}{\cal O}^{\L_R\L_0}_{g~2}|_0=0~~;~~
{\cal O}^{\L_R\L_0}_{g~4}|_0=1
\ena

\bea\label{rcomm}
{\cal O}^{\L=0\L_0}_{m^2~2}(0;0)=0~~;~~
\partial_{m^2}{\cal O}^{\L_R\L_0}_{m^2~2}|_0=1~~;~~
\partial_{p^2}{\cal O}^{\L_R\L_0}_{m^2~2}|_0=0~~;~~
{\cal O}^{\L_R\L_0}_{m^2~4}|_0=0
\ena
Observe that ${\cal O}^{\L\L_0}_{m^2}$ is a functional insertion 
of the form (\ref{fluxO2}) which
at $\L_0$ is the bare term 
${\cal O}^{\L_0}_{m^2}=m^2 B({\L_0 \over \L_R})\Phi^2$, with $B^{(0)}=1$.
Using the Quantum Action Principle we can express ${\cal O}_k$ and
${\cal O}_{m^2}$ in terms of
$\Phi {\delta \over \delta \Phi }$ and $m^2 {\partial \over
\partial m^2 }$ acting on a functional.
In the case of the number operator eqs. (\ref{N1}, \ref{N}, \ref{N2})
remain true, provided $D$ is the massive propagator.
From (\ref{N2}) we get
\bey\label{rcNm}
&&{\cal O}^{\L=0\L_0}_{N~2}(0;0)=0~~~~~;~~~~~~~
\partial_{p^2}|_{0} {\cal O}_{N~2}^{\L_R\L_0}(0;p)=2+\alpha_2~~
\nonumber \\
&&\partial_{m^2}|_{0} {\cal O}_{N~2}^{\L_R\L_0}(m;0)=2+\alpha_1~~;~~
{\cal O}^{\L_R\L_0}_{N~4}(0;0,0,0)=4g+\alpha_3
\eny
with
\bea\label{am}
\alpha_1 = \hbar \partial_{m^2}|_{0} \int_q K_{\L_R}(m;q)
D_{\L_R\L_0}(m;q) L^{\L_R\L_0}_4(m;q,-q,0)
\ena
while $\alpha_2$ and $\alpha_3$ are the same as in eq.(\ref{aaa}).
Eq.(\ref{N2}) shows that ${\cal O}_{N}$ satisfies eq.(\ref{bob}) and
then it can be considered as a mass-independent operator of dimension four.

Using eqs.(\ref{rcokm}- \ref{rcNm}) we can write
\bea\label{Nm4}
{\cal O}^{\L\L_0}_{N}[\Phi] = (2+\alpha_1){\cal O}^{\L\L_0}_{m^2}[\Phi] +
(2+\alpha_2){\cal O}^{\L\L_0}_k[\Phi]
+(4g + \alpha_3) {\cal O}^{\L\L_0}_g[\Phi] + {\cal E}^{\L\L_0}[\Phi]
\ena
where ${\cal E}^{\L\L_0}$ is an evanescent flowing functional.
Indeed ${\cal E}$ defined in (\ref{Nm4}) is a functional which
satisfies the flow equation (\ref{fluxO}), it has vanishing
mass-independent renormalization conditions and it fulfils the
condition (\ref{bob}) for $D=4$;  ${\cal E}$ is not identically zero due to the
remainder of irrelevant terms in  ${\cal O}_N$. 
  
To complete our proof of the Quantum Action Principle we now consider
\bea\label{MMMM}
{\cal O}^{0\infty}_{M^2}[DJ] =
exp \{-{1\over \hbar} \omega^{0\infty}[DJ;m^2]~\}\hbar
m^2\partial_{m^2} exp \{{1\over \hbar} \omega^{0\infty}[DJ;m^2]~\}
\ena
where the function $\omega^{0\L_0}$ was introduced in the massless
case in eq.(\ref{om}). 
For finite $\L_0$ we define (with some arbitrariness) a functional
which in the limit $\L_0 \to \infty$ gives the functional defined in 
eq.(\ref{MMMM}); thus we define
\bea\label{conn}
{\cal O}^{0\L_0}_{M^2}[\Phi] =
e^{{1\over \hbar}L^{0\L_0}}
[ -{m^2 \over 2}\Phi^2 +\hbar m^2 \partial_{m^2} - \hbar \Phi m^2 D 
{\delta \over \delta \Phi}] e^{-{1\over \hbar}L^{0\L_0}[\Phi,m^2]}
\ena
Note that in these two last formulas, differently from the previous
one, $ \partial_{m^2}$ does not act on the external legs $\Phi =
D^{0\L_0}J$.
The corresponding flowing functional is
\bey\label{MM}
&&{\cal O}^{\L\L_0}_{M^2}[\Phi] = 
e^{{1\over \hbar}L^{\L\L_0} [\Phi,m^2]} e^{-\hbar \Delta^{0\L}}
[ -{m^2 \over 2}\Phi^2 +\hbar m^2 \partial_{m^2} - \hbar \Phi m^2 D 
{\delta \over \delta \Phi}] e^{-{1\over \hbar}L^{0\L_0}[\Phi,m^2]} \nonumber \\
&&= -{m^2 \over 2}\Phi^2 - m^2 \partial_{m^2}L_{\L\L_0} +
m^2\Phi (1-K_{\L})D L_{\L\L_0}' + \\
&&{m^2 \over 2}L_{\L\L_0}'
D[D (1-K_{\L})K_{\L}+ (\partial_{m^2}K_{\L})]L_{\L\L_0}' 
 -{m^2 \over 2} \hbar tr D[D (1-K_{\L})K_{\L}+
(\partial_{m^2}K_{\L})]L_{\L\L_0}'' \nonumber
\eny
${\cal O}^{\L_0}_{M^2}$ satisfies the condition (\ref{bob}) for
$D=4$.
From eq.(\ref{MM}) we see that ${\cal O}^{\L\L_0}_{M^2}$ satisfies 
the renormalization condition (\ref{O22}) and
\bey
&&\partial_{m^2}|_{0}{\cal O}^{\L_R\L_0}_{M^2 2}(m;0) = -1 -
\alpha_m  \nonumber \\
&&\alpha_m \equiv {\hbar \over 2} \int_q D(q)[D(1-K_{\L_R})K_{\L_R}+
 {K_{\L_R}' \over \L_R^2}](q) L^{\L_R\L_0}_4(0;q,-q,0)
\eny
the parameters in the other renormalization 
conditions for an operator of dimension four being zero. 
In the above equation $K_{\L_R}'(p)$ is the derivative of the
function $K$ with respect to its argument.
We can then write
\bea\label{Mm}
{\cal O}^{\L\L_0}_{M^2}[\Phi, m^2] = -(1+\alpha_m){\cal O}^{\L\L_0}_{m^2}
+{\cal E}^{\L\L_0}
\ena
where ${\cal E}^{\L\L_0}$ in a new evanescent functional.
Now from (\ref{ergm}), (\ref{Nm4}) and (\ref{Mm}) we obtain
\bey\label{ergmm}
( \L_R {\partial \over \partial \L_R } +\beta {\partial \over \partial g } )
L^{\L\L_0} - \gamma_{\phi}{\cal O}^{\L\L_0}_N
+ \gamma_m {\cal O}^{\L\L_0}_{M^2}  = {\cal E}^{\L\L_0}
\eny
where
\bey\label{bega}
\beta = \beta_3 + \gamma_{\phi} (4g+\alpha_3) ~~;~~
\gamma_{\phi} = {-\beta_2 \over 2+\alpha_2} ~~;~~
\gamma_m =- {\beta_1 + \gamma_{\phi} (2+\alpha_1) \over 1 + \alpha_m}
\eny
$\beta,\gamma_{\phi}$ and $\gamma_m$ in (\ref{bega}) are independent
from $m^2$ so that, for $\L_0 \to \infty$, they become functions of
$g$ only and coincide with those in (\ref{beta}). The operator
insertion 
 $ {\cal E}^{\L\L_0}$ is evanescent and, for  $\L_0 \to
\infty$ and $\L \to 0$, one gets from (\ref{N2},~\ref{MMMM}) and
(\ref{ergmm}) 
the mass-independent renormalization group equation  (\ref{rge}).

In the appendix we will show that the last term in the left hand side
of (\ref{ergmm}) is vanishing in the limit $(\L,m) \to 0$ and
 we get eq.(\ref{rgez}) from  eq.(\ref{ergmm}).

Let us discuss the relation between the mass-independent
renormalization group equation and the Callan-Symanzik equation \cite{CS}.
Note that eq.(\ref{Mm}) gives a simple relation between the
differentiation with respect to the mass and the insertion of the
composite operator $m^2\Phi^2$ acting on the Green functions.
In eq.(\ref{rcomm}) this operator was renormalized, in our
mass-independent scheme, as an operator of dimension four, but it is
trivial to check that
\bea\label{N42}
{\cal O}^{\L\L_0}_{m^2}[\Phi] = m^2 {\cal O}^{\L\L_0}_{\phi^2}[\Phi]
\ena
where ${\cal O}^{\L\L_0}_{\phi^2}[\Phi]$ satisfies  eq.(\ref{bob}) for
$D=2$ and the renormalization condition
\bey
 {\cal O}^{\L_R\L_0(l)}_{\phi^2~2}(0;0)= \delta^{l,0} \nonumber
\eny
Observe that ${\cal O}^{\L\L_0}_{\phi^2}[\Phi]$ is infrared singular,
in the sense that it does not exist in the limit $(\L,m) \to (0,0)$.

In the language of Zimmermann normal product \cite{Zimm} 
(at scale $\L_R$ and in our scheme) eq.(\ref{N42}) could be written as
\bea\label{Zim}
N_4^{\L_R}[m^2\Phi^2] = m^2 N_2^{\L_R}[\Phi^2]
\ena
The Callan-Symanzik equation can be obtained by
taking a linear combination of $ {\cal O}_g,~ {\cal O}_N,~ {\cal
O}_{\L_R}$ and $ {\cal O}_{M^2}$ in such a way that their sum is equal
to an operator of dimension two and the scaling operator
$\L_R \partial_{\L_R}+m\partial_m$ is obtained. As a consequence of 
 eq.(\ref{N42}) it has the form
\bey
(\L_R {\partial \over \partial \L_R }+ m{\partial \over \partial m}+
\beta {\partial \over \partial g } + 
\gamma_{\phi} J{\delta \over \delta J } ) W[J] = 
m^2 \sigma {\cal O}^{0\infty}_{\phi^2}[DJ] 
\eny
where the beta and gamma functions are the same as in eq.(\ref{bega}) and
\bey
\sigma = -(2+\gamma_m)(1+\alpha_m)
\eny

We conclude making an observation on the
mass-independent renormalization conditions (\ref{renLm})
in which  $\rho_1$ is not tuned according to  (\ref{ren2m}) but it is
simply set to zero and
 $\tilde{\rho}_1=\rho_2=0$ and $\rho_3=g$ as before.
The proof of ultraviolet renormalization is unaffected, so that
$L_n^{0\infty}(m;p)$ exists for $m >0$; however the limit for $m \to 0$
is singular. The bare lagrangian (\ref{barem}) has still a polynomial
dependence on $m$,  the renormalization group equation  (\ref{rge})
holds, with the same formulae  (\ref{beta}) for $\beta(g)$ and 
$\gamma_{\phi}(g)$,
while $\gamma_m(g,m)$ has a simple dependence on the mass:
$\gamma_m =\gamma_{m,1}(g)+{\L_R^2 \over m^2}\gamma_{m,2}(g)$.
The singularity of its limit for $m \to 0$ reflects the corresponding
singularity of the Green functions $L_n^{0\infty}(m;p)$ in this limit.

\section{Conclusion}

In the Wilson-Polchinski approach to renormalization it is natural to
set renormalization conditions at zero momentum at a Wilsonian scale
$\L_R > 0$, even in massless theories.
This renormalization scheme is similar to the hard-soft (HS)
renormalization scheme proposed in \cite{LM}
 to renormalize in a simple way massless theories with BPHZ, 
though differing from it
in the choice of the cut-off function separating soft and hard modes.

In this paper we have shown that the renormalization group equation in 
a Wilsonian HS scheme in massless and in massive
$g\phi^4$ can be deduced from the exact renormalization flow and the
use of the Quantum Action Principle; at scale $\L$ one obtains an
effective renormalization group equation, which reduces to the usual
renormalization group equation at $\L = 0$.
We give exact formulae for the beta and gamma functions in terms of
the Wilsonian vertices at scale $\L_R$ and we verify that their lowest
order coefficients have the expected values \cite{Gross}.
In the massive case we have 
introduced a mass-independent hard-soft renormalization scheme, and we
have proven its consistency. The beta function and $\gamma_{\phi}$ 
are the same as in the
massless case. For non-exceptional momenta the massless limit of the
massive Green functions exists and it coincides to the massless Green
functions.

The relevant terms of the massive theory are given by the Maclaurin
expansion in momenta and in the mass of the Wilsonian effective action
at scale $\L_R$ , up to dimension four.
In this scheme the renormalization of the massless theory is the
particular case $m=0$ of the massive theory and its
 relevant terms are simply given by the $m=0$ part of the above mentioned
 Maclaurin expansion.

It should be possible to generalize our results to other field
theories, in particular to gauge theories, using HS renormalization
conditions compatible with the effective Ward identities
\cite{Becchi2}.
It is obvious that in these theories one can compute the beta and gamma
functions in terms of the Wilsonian vertices at scale $\L_R$.
It is an interesting question whether the mass-independent Wilsonian 
HS scheme introduced here can be generalized to any renormalizable
theory involving mass parameters.

\vskip 2cm

\centerline{\bf{Appendix A: Renormalization.}}
\vskip 1cm

In this Appendix we prove the ultraviolet finiteness of the theory in
the mass-independent scheme; we will consider generic renormalization
conditions  (\ref{renLm}), that is we will not need to choose a
particular value for $\rho_1$ to prove the convergence of the
effective Green functions at scale $\L \geq \L_R$.
 This will be done in two steps.
We start by proving that eq.(\ref{renLm}) implies the bounds
(\ref{rt}); to this end we write down  explicitely the flow equation 
for the $n$-point Green functions at $l$ loops:
\bey\label{Ap1}
&&\partial_{\L} L_n^{\L\L_0 (l)}(m;p_1,...,p_{n-1}) = {1\over 2} \int_q 
\partial_{\L} D_{\L\L_0}(m;p)
 L_{n+2}^{\L\L_0 (l-1)}(m;q,-q,p_1,...,p_{n-1})+ \nonumber \\ 
&&-{1\over 2} \sum_{n_1+n_2=n+2}  L_{n_1}^{\L\L_0 (l_1)}(m;p_{i_1},...,p_{i_{n_1-1}})
\partial_{\L} D_{\L\L_0}(m;P)
 L_{n_2}^{\L\L_0 (l_2)}(m;P,p_{i_{n_1}},...,p_{i_{n-1}})
\eny
where $P \equiv - \sum_1^{n_1-1}p_{i_k}$ and $l_1+l_2=l$; in the sum
there is a symmetrization with respect to $p_1,...,p_{n-1}$ and 
$- \sum_1^{n-1}p_i$ .
In the proof a key role is played by the following bound on the propagator
\bea\label{Ap2}
|| \partial_{m^2}^r \partial_{p}^z\partial_{\L} D_{\L\L_0}||_{\L} < 
C \L^{-3-z-2r}
\ena
where the norm was introduced in (\ref{norM}). Here and in the
following $C$ will denote constants whose value is not important.
Applying $\partial_{m^2}\partial_{p}^z$ to both sides of  (\ref{Ap1}),
using  (\ref{Ap2}) one easily arrives at
\bey\label{Ap3}
&&||\partial_{\L} \partial_{m^2}^r \partial_{p}^z L^{\L\L_0 (l)}_n||_{\L} < 
C_1 \sum_{r_1+r_2=r}\L^{1-2r_1}
|| \partial_{m^2}^{r_2} \partial_{p}^z L^{\L\L_0 (l-1)}_{n+2}||_{\L}
\nonumber \\
&&+ C_2 \sum  \L^{-3-z_3-2r_3}
|| \partial_{m^2}^{r_1} \partial_{p}^{z_1} L^{\L\L_0 (l_1)}_{n_1}||_{\L} 
|| \partial_{m^2}^{r_2} \partial_{p}^{z_2} L^{\L\L_0 (l_2)}_{n_2}||_{\L} 
\eny
where in the second term the sum is also over $\sum r_i =r$ and
$\sum z_i = z$.
The proof is made by induction in the loop index $l$ and in the $l$-th
step we will use an induction in the index $n$ of the number of legs.
In general eq.(\ref{rt}) will be proven first in the case $w=1$ and
then, with a suitable integration, for $w=0$. 
From  (\ref{Ap1}), which for $l=0$ has not the first term, one gets
$L_2^{\L\L_0 (0)}=0$ and $L_4^{\L\L_0 (0)}=g=\rho_3^{(0)}$, which
fulfils trivially  eq.(\ref{rt}). Let us consider now $n>4$: in the
right hand side, because of $L_2^{\L\L_0 (0)}=0$, only the terms with 
 $L_{n'}^{\L\L_0 (0)}$ with $n'<n$ are involved, so that we use an
induction in $n$, and we arrive at
\bea
||\partial_{\L} \partial_{m^2}^r \partial_{p}^z L_n^{\L\L_0(0)}||_{\L} < 
C \L^{3-n-z-2r} \nonumber
\ena
From this, integrating from $\L=\L_0$,
\bea
|| \partial_{m^2}^r \partial_{p}^z L_n^{\L\L_0(0)}||_{\L} \leq\int_{\L}^{\L_0}
d\L'
||\partial_{\L'} \partial_{m^2}^r \partial_{p}^z L_n^{\L'\L_0(0)}||_{\L'} \leq 
C \L^{4-n-z-2r} \nonumber
\ena
For $l>0$ we assume eq.(\ref{rt}), for every $z,r,w$ and $l' < l$.
 We start considering the case $n=2$;
 only $L_{n'}^{(l')}$ with  $l' < l$ are present in the
right hand side of  eq.(\ref{Ap3}) so that, from the induction
hypothesis, we get
\bea
||\partial_{\L} \partial_{m^2}^r \partial_{p}^z L_2^{\L\L_0 (l)}||_{\L} \leq 
 \L^{1-z-2r} P(ln({\L \over \L_R})) \nonumber
\ena
Now for $2r+z > 2$ we integrate in $\L$ from $\L=\L_0$ arriving at 
 eq.(\ref{rt}) as shown in the case $l=0$. Consider now the case
$2r+z =2$. Using  eq.(\ref{renLm}) we can write
\bey
&&\partial_{m^2}^r \partial_{p}^z L_2^{\L\L_0 (l)}(m;p) =\nonumber \\
&& \delta_{r,1} 
\tilde{\rho}_1^{(l)} + \delta_{r,0}\rho_2^{(l)}
+ \int_0^p dp' \partial_{m^2}^r \partial_{p'}^{z+1}
L_2^{\L\L_0 (l)}(0;p')
+ \int_0^{m^2} dm'^2 \partial_{m'^2}^{r+1} \partial_{p}^z
L_2^{\L\L_0 (l)}(m';p)
\eny
In the integrands we can use  eq.(\ref{rt}) for $2r+z > 2$
so that we arrive again at   eq.(\ref{rt}) for $2r+z = 2$.

At this point analogous considerations can be applied to the case
$n=2,r=z=0$, where now the renormalization condition on
$L_2^{\L\L_0 (l)}(0;0)$ plays the fondamental role.

For $n=4$ we observe that
the only addendum in the right hand side of eq.(\ref{Ap3}) which
contains terms with $l$ loops is 
$||\partial_{m^2}^r \partial_{p}^zL_2^{\L\L_0(l)}||$ for which we have just
now proven the bound, therefore 
$||\partial_{\L} \partial_{m^2}^r \partial_{p}^z L_4^{\L\L_0 (l)}||_{\L}
\leq \L^{-1-z-2r} P(ln({\L \over \L_R}))$. 
From this it follows that
$|| \partial_{m^2}^r \partial_{p}^z L_4^{\L\L_0 (l)}||_{\L}
\leq \L^{-z-2r} P(ln({\L \over \L_R}))$, which 
in the case $z+2r > 0$ it is obtained using
integration from $\L = \L_0$, while for $z=r=0$ it is obtained using the Taylor
reconstruction formula:
\bey\label{TAY}
L^{\L\L_0 (l)}_4(m;p_1,p_2,p_3)=&& \rho_3^{(l)} + \int_{\L_R}^{\L} d \L'
\partial_{\L'}L^{\L'\L_0 (l)}_4(0;0)
+ \int_0^{m^2}dm'^2 {\partial \over \partial m'^2} L^{\L\L_0
(l)}_4(m';0) \nonumber \\
&&+ \int_0^1dt \sum p_i^{\mu}  {\partial \over \partial q_i^{\mu}} L^{\L\L_0
(l)}_4(m;q_i)|_{q_i=tp_i}
\eny
For $n > 4$ one realizes, as in the previous case, that in the right
hand side of eq.({\ref{Ap3}) are contained only $L_{n'}^{(l')}$ for
$l'<l$ or, if $l'=l$, with $n'<n$; thus we arrive at
$||\partial_{\L} \partial_{m^2}^r \partial_{p}^z L_n^{\L\L_0 (l)}||_{\L}
\leq \L^{3-n-z-2r} P(ln({\L \over \L_R}))$ and then, integrating from
$\L=\L_0$, at eq.(\ref{rt}) also for $w=0$.

The second step consists in proving eq.(\ref{BOU}).
To this aim let us differentiate with respect to $\L_0$ both sides of
eq.(\ref{Ap1}). The resulting equation 
\bey\label{Ap4}
&&\partial_{\L}{\cal{O}}_n^{\L\L_0 (l)}(p_1,...,p_{n-1}) =
{1\over 2}\int_q \partial_{\L} D_{\L}(m;q) 
{\cal{O}}_{n+2}^{\L\L_0 (l-1)}(m;q,-q;p_1,...,p_{n-1}) - \nonumber \\
&&\sum L_{n_1}^{(l_1)}(m;p_{i_1},...,p_{i_{n_1-1}}) \partial_{\L}D_{\L}(P)
{\cal{O}}_{n_2}^{\L\L_0 (l_2)}(m;P,p_{i_{n_1}},...,p_{i_{n-1}})
\eny
coincides with the flow equation of the operator insertion (\ref{fluxO}) with 
${\cal{O}}_n^{\L\L_0} \equiv \partial_{\L_0}L_n^{\L\L_0}$.
Now the boundary conditions on ${\cal{O}}^{\L\L_0}$   from its definition
are, for $n+2r+z \geq 5$ using (\ref{rt})
\bea\label{Ap5}
||\partial_{m^2}^r \partial_{p}^z {\cal{O}}_n^{\L_0(l)}||_{\L_0} =
||\partial_{\L}|_{\L=\L_0} \partial_{m^2}^r \partial_{p}^z 
L_n^{\L\L_0(l)}||_{\L_0} \leq \L_0^{3-n-z-2r} P(ln({\L_0 \over \L_R}))
\ena
while the renormalization conditions, as operator of dimension $4$,
are vanishing.

From eq.(\ref{Ap4}) one easily gets, using eq.(\ref{rt}),
\bey\label{Ap6}
&&||\partial_{\L}\partial_{m^2}^r \partial_{p}^z
 {\cal{O}}_n^{\L\L_0(l)}||_{\L} \leq
C_1\sum \L^{1-2r_1} 
||\partial_{m^2}^{r_2} \partial_{p}^z {\cal{O}}_{n+2}^{\L\L_0(l-1)}||_{\L}
\nonumber \\
&&+C_2 \sum  \L^{4-n_1-z_1-2r_1} P(ln({\L \over \L_R}))
\L^{-3-z_2-2r_2}
||\partial_{m^2}^{r_3} \partial_{p}^{z_3}
 {\cal{O}}_{n_2}^{\L\L_0(l_2)}||_{\L}
\eny
Eq.  (\ref{BOU}) is proven by induction in $l$.
For $l=0$ one has 
${\cal{O}}_2^{\L\L_0(0)}={\cal{O}}_4^{\L\L_0(0)}=0$, and for $n \geq 5$ 
one uses an
induction in $n$ as in the first theorem. Analogously for $l> 0$
starting from $n=2$, $2r+z> 2$, then $2r+z=2$ and  $2r+z=0$ using
 vanishing renormalization conditions.
Subsequently we go to the case $n=4,~ 2r+z >0$ and then, using the
 renormalization conditions, to the case  $n=4, 2r+z =0$. For the
cases with
$n\geq 5$, using the induction hypothesis one has
\bea
||\partial_{\L}\partial_{m^2}^r \partial_{p}^z
 {\cal{O}}_n^{\L\L_0(l)}||_{\L} \leq
{\L^{4-n-2r-z} \over \L_0^2} P(ln({\L_0 \over \L_R})) \nonumber
\ena
and then integrating from $\L=\L_0$ one arrives at the bound  (\ref{BOU}).

Note that if the renormalization conditions are not vanishing but are
consistent with the behaviour $\L_0^{-2}$ for $\L_0 \to \infty$ we
still  arrive  at eq.  (\ref{BOU}); with few modifications one can
prove that if the flow of a composite operator of dimension $D$
satisfies 
\bea
||\partial_{m^2}^r \partial_{p}^z {\cal{O}}_n^{\L_0}||_{\L_0} \leq
\L_0^{D-n-2r-z} P(ln({\L_0 \over \L_R}))
\nonumber
\ena
and their  renormalization conditions go to zero at
least as $\L_0^{-1}$, then
\bea
||\partial_{m^2}^r \partial_{p}^z {\cal{O}}_n^{\L\L_0}||_{\L} \leq
{\L^{D+1-n-2r-z} \over \L_0} P(ln({\L_0 \over \L_R})) \to 0
\nonumber
\ena
for $\L_0 \to \infty$.

\vskip 2cm

\centerline{\bf{Appendix B: Infrared finiteness.}}
\vskip 1cm

In this appendix we want to prove the infrared regularity of the
mass-independent scheme. The last goal will be to prove that with
suitable renormalization conditions on $L_2^{\L_R\L_0}(0;0)$ - arbitrary
in Appendix A - the vertices $L_n^{\L\L_0}(m;p_1,...,p_{n-1})$ with
non-exceptional momenta converge for $(m,\L,\L_0) \to (0,0,\infty)$.

The first step consists in proving the bounds (\ref{bs}).
In this appendix we will pay more attention to the dependence on the
mass $m$. We begin by noticing that
\bey\label{ApB1}
|\partial_{\L} \partial_{m^2}^r \partial_p^z D_{\L\L_0}(m;p)| \leq &&
C(\L+ {m \over 2})^{-3-z-2r}\theta (4\L^2 -m^2-p^2) \nonumber \\
 \leq && C(\L+ {m \over 2})^{-3-z-2r}\theta (\L -{m\over 2})
\eny
Moreover from the flow equation, for $\L \leq \eta_1/2$
\bey\label{ApB2}
&&||\partial_{\L} \partial_{m^2}^r \partial_p^{z,z'} L^{\L\L_0 (l)}_n
||_{\L}^k \leq C_1 \sum (\L+ {m \over 2})^{1-2r_1} 
|| \partial_{m^2}^{r_2} \partial_p^{z,z'} L^{\L\L_0 (l-1)}_{n+2}
||_{\L}^{k+2}
\nonumber \\
&&+C_2 \sum_{n_1 \leq k+1} (\L+ {m \over 2})^{-3-z_3-2r_3} 
|| \partial_{m^2}^{r_1} \partial_p^{z_1} L^{\L\L_0 (l_1)}_{n_1}
||_{\L}^{n_1-1}
|| \partial_{m^2}^{r_2} \partial_p^{z_2,z'} L^{\L\L_0 (l_2)}_{n_2}
||_{\L}^{k-n_1+2}
\eny
where $z_1+z_2+z_3=z$ and
where for the second term in the right hand side of  (\ref{ApB2}) we have
used the fact that, because of the first line of  (\ref{ApB1}), the
terms in which 
$||\partial_{m^2}^{r_1} \partial_p^{z_1}
L_{n_1}^{\L\L_0(l_1)}||_{\L}^{k_1}$ 
should appear, with $k_1<n_1-1$, are vanishing for  $\L \leq \eta_1/2$.

Consider first the case $l=0$.
For $n=2$ and $n=4$ all the bounds are trivially satisfied. For 
 $n>4$ the induction in $n$ works. If $3 \leq k < n-1$ we get 
$||\partial_{\L} \partial_{m^2}^r \partial_p^{z,z'} L_n^{\L\L_0 (0)}||_{\L}^k
\leq \L^{-1-2[{k-1\over 2}]-z-2r}$,
and then integrating from $\L=\L_R$:
\bey
||\partial_{m^2}^r \partial_p^{z,z'} L^{\L\L_0 (0)}_n
||_{\L}^k \leq &&
\int_{\L}^{\L_R} d\L'
||\partial_{\L'} \partial_{m^2}^r \partial_p^{z,z'} L^{\L'\L_0 (0)}_n
||_{\L'}^k 
+|| \partial_{m^2}^r \partial_p^{z,z'} L^{\L_R\L_0 (0)}_n
||_{\L_R}^k  \nonumber \\
\leq &&C (\L+ {m \over 2})^{-2[{k-1\over 2}]-2r-z}
\nonumber
\eny
A similar discussion holds for the case $k=n-1$.
If $k< 3$ the right hand side of eq.(\ref{ApB2}) is vanishing, thus
$||\partial_{\L} \partial_{m^2}^r \partial_p^{z'} L_n^{\L\L_0 (0)}||_{\L}^{k} =0$ and
$|| \partial_{m^2}^r \partial_p^{z'} L_n^{\L\L_0 (0)}||_{\L}^{k} \leq C$.

Let us consider now $l >0$. The induction hypothesis is  eq.(\ref{bs})
for $l'<l$; the $l$-th step of the proof is made by induction in
$n$. For $n=2$ the  right hand side of  (\ref{ApB2}) contains only
terms with loop index $l'<l$; using the  induction hypothesis we get
the bound of eq.(\ref{bs}) for $w=1$, $k=0$ and  $k=1$. Notice that
for $k=0$ the second term of eq.(\ref{ApB2}) is absent, and that
integrating from $\L=\L_R$ we get (\ref{bs}) also for $k=0=w$.
Now consider $k=1$.
 If $2r+z \geq 2$, integrating from
$\L = \L_R$ we get the result also for $w=0$. For the case $r=z=0$ we
notice that since 
$|\partial_{\L} L_2^{\L\L_0(l)}(0;0)| < \L P(ln {\L_R \over \L})$
this function is integrable in a neighborhood of $\L =0$. Thus 
\bey
 L^{\L\L_0 (l)}_2(0;0) = \rho_1^{(l)} - \int_{\L}^{\L_R} d\L'
\partial_{\L'} L^{\L'\L_0 (l)}_2(0;0)
\nonumber
\eny
has a finite limit for $\L \to 0$ and since the second term does not
depend on $\rho_1^{(l)}$, this constant can be chosen such that
$\lim_{\L \to 0} L_2^{\L \L_0(l)}(0;0) = 0$. With this choice
\bey
| L^{\L\L_0 (l)}_2(0;0)| \leq  \int_{0}^{\L} d\L'
|\partial_{\L'} L^{\L'\L_0 (l)}_2(0;0)| \leq \L^2
P(ln {\L_R \over \L}) \leq  
(\L+ {m \over 2})^2 P(ln {\L_R+ {m\over 2} \over \L + {m\over 2}})
 \nonumber
\eny
Thus from
\bey
 L^{\L\L_0 (l)}_2(m;p) =  L^{\L\L_0 (l)}_2(0;0) +
\int_0^p dp'\partial_{p'}  L^{\L\L_0 (l)}_2(m;p')
+\int_0^{m^2} dm'^2\partial_{m'^2} L^{\L\L_0 (l)}_2(m';p) 
\nonumber
\eny
one gets
\bey
|| L^{\L\L_0 (l)}_2||_{\L}^1 \leq 
(\L+ {m \over 2})^2 P(ln {\L_R+ {m\over 2} \over \L + {m\over 2}})
\nonumber
\eny

For $n > 2$ in the right hand side of (\ref{ApB2}) there are only terms
$L_{n'}^{(l')}$ with $l'<l$ or, if $l'=l$, with $n'<n$.

For $k\geq 1$ every term in  (\ref{ApB2}) has at least one `small
momentum'. From the induction hypothesis we arrive at the bound
(\ref{bs}) for $w=1$ and then, integrating from $\L=\L_R$, at the
bound for $w=0$.  If $k=0$ only the first term in  (\ref{ApB2}) has to
be taken into account and thus
\bey
||\partial_{\L} \partial_{m^2}^r \partial_p^{z'} L_n^{\L\L_0(l)}||^0 
\leq (\L+ {m \over 2})^{1-2r} P(ln {\L_R+ {m\over 2} \over \L + {m\over 2}})
\nonumber
\eny
and then one gets easily eq.(\ref{bs}) also for $w=0$.

At this point of the proof one could easily conclude that, after
taking the limit $\L_0 \to \infty$, the infrared part of the proof
could be done starting from the renormalization point
$L^{\L_R\infty}[\Phi]$, showing that the limit $\L \to 0$ exists, and
moreover that the limit $\L \to 0,~ m \to 0$ exists, leading to Green
functions which are $C^{\infty}$ functions for non-exceptional
momenta.
But actually it is possible to prove that the global limit
$(\L,m,\L_0) \to (0,0,\infty)$ exists, so that no ambiguities in the
definition of the theory are involved. Let us outline this proof.
To this aim one should generalize the second theorem of 
Appendix A and prove that, if $\rho_1^{(l)}$ are chosen loop by loop
according to eq.(\ref{ren2m}), then the quantities
$\partial_{\L_0} L_n^{\L\L_0(l)}$ are such that
$||\partial_{\L}^w \partial_{m^2}^r \partial_p^{z,z'} \partial_{\L_0}
L_n^{\L\L_0(l)}||_{\L}^k$
satisfy the bound of eq.(\ref{bs}), multiplied by 
$P(ln {\L_0 \over \L_R} )/ \L_0^2$.

These bounds are essentially obvious, and are proven by induction;
we do not reproduce the proof in detail. They hold because: 
i) $\partial_{\L_0} L_n^{\L_R\L_0(l)}(p_1,..,p_{n-1})$ satisfies
eq.(\ref{BOU});
ii) the flow equation  (\ref{fluxO}) is linear; iii) loop by loop the key
renormalization condition on the positive dimension Green function is
satisfied. Indeed 
\bey
\lim_{\L \to 0} \partial_{\L_0} L_2^{\L\L_0(l)}(0;0) =
\lim_{\L \to 0} \partial_{\L_0} \int_0^{\L} d\L'  \partial_{\L'}
 L_2^{\L'\L_0(l)}(0;0)= \lim_{\L \to 0} \int_0^{\L} d\L' 
\partial_{\L'}  \partial_{\L_0} L_2^{\L'\L_0(l)}(0;0) = 0
\nonumber
\eny
where in the second and third equality we have used the 
induction hypothesis in 
eq.(\ref{fluxO}), for which the last integrand is bounded by 
${\L \over \L_0^2} P(ln {\L_0 \over \L_R}) P(ln {\L_R \over \L})$. 
Using this bound in the third addendum of eq.(\ref{tz}), as well as
those of the first theorem in Appendix B for the first two terms, one
arrives easily at the assertion
\bey
&&| \partial_p^{z'} L_n^{\L_1\L_0(l)}(m_1;p) -
 \partial_p^{z'} L_n^{\L_2\L_0'(l)}(m_2;p)| \leq
\int_{\L_1}^{\L_2}d\L' ||\partial_{\L'}\partial_p^{z'}
L_n^{\L'\L_0'(l)}(m_2)||^0 \nonumber \\
&&+ \int_{m_1^2}^{m_2^2}dm'^2 ||\partial_p^{z'} \partial_{m'^2}
L_n^{\L_1\L_0'(l)}(m')||^0
+ \int_{\L_0}^{\L_0'} d\L_0''  ||\partial_{\L_0''}
\partial_p^{z'} L_n^{\L_1\L_0''(l)}(m_1)||^0 \to 0
\nonumber
\eny
for $(\L_{1,2},m_{1,2},\L_0,\L_0') \to (0,0,\infty,\infty)$.

As a last remark let us notice that from the bound  (\ref{bs}) for
$k=w=0$ and $r=1$ it follows that
$\partial_{m^2}L_n^{0\L_0(l)}(m;p_1,...,p_{n-1})$ 
has only logarithmic divergences in $m$ for $m \to 0$ and therefore 
$m^2\partial_{m^2}L_n^{0\L_0(l)}(m;p_1,...,p_{n-1}) \to 0$ 
for non-exceptional momenta in the limit 
$m \to 0$, so that the Gell-Mann and Low renormalization group
equation in the massless case can be obtained as the $m \to 0$ limit
of the corresponding equation for the massive case.

\vskip 2cm

\centerline{\bf{Appendix C: Exact renormalization group equation.}}
\vskip 1cm

In this appendix we examine some points regarding the proof of the
exact renormalization group equation (\ref{ERG}); similar
considerations can be repeated for eq.(\ref{ERGm}).

Because of eq.(\ref{rel}), the functional $ L^{\L_R\L_0(l)}[\Phi]$
will depend on $2l+1$ arbitrary parameters, that is
$(\rho_2^{(1)},...,\rho_2^{(l)}) =\underline{\rho}_2^{(l)}$ and
$(\rho_3^{(o)},...,\rho_2^{(l)}) =\underline{\rho}_3^{(l)}$,
and we will use the notation 
$L^{\L\L_0(l)}[\L_R,\underline{\rho}_2^{\L_R\L_0},
\underline{\rho}_3^{\L_R\L_0};\Phi]$
for this functional.
The standard values of $\underline{\rho}_2^{(l)}$ and 
$\underline{\rho}_3^{(l)}$
are determined at $\L=\L_R$ in eq.(\ref{rc}).
In perturbation theory it is true that
\bea\label{ERR}
L^{\L\L_0(l)}[\L_R,\underline{\rho}_2^{\L_R\L_0},
\underline{\rho}_3^{\L_R\L_0};\Phi]=
L^{\L\L_0(l)}[\L_R',\underline{\rho}_2^{\L_R'\L_0},
\underline{\rho}_3^{\L_R'\L_0};\Phi]
\ena
since the relation between the bare coefficients  and the parameters
appearing in the renormalization conditions is invertible.

Differentiating this equation with respect to $\L_R'$ in $\L_R'=\L_R$
we get
\bea\label{ERGG}
[\L_R {\partial \over \partial \L_R }
+ \sum_{l'=1}^l \beta_2^{(l')} 
{\partial \over \partial \rho_2^{(l')} }
 + \sum_{l'=0}^l \beta_3^{(l')}{\partial \over \partial  \rho_3^{(l')} }] 
 L^{\L\L_0(l)}[\L_R,\underline{\rho}_2,\underline{\rho}_3;\Phi]= 0
\ena
The functionals
\bey\label{okg}
{\cal O}_{k,l'}^{\L\L_0(l)}[\Phi] \equiv
{\partial \over \partial \rho_2^{(l')} }
 L^{\L\L_0(l)}[\L_R,\underline{\rho}_2,\underline{\rho}_3;\Phi]
~~;~~{\cal O}_{g,l'}^{\L\L_0(l)}[\Phi] \equiv
{\partial \over \partial \rho_3^{(l')} }
 L^{\L\L_0(l)}[\L_R,\underline{\rho}_2,\underline{\rho}_3;\Phi]
\eny
satisfy eq.(\ref{fluxO}) and in fact define integrated composite
operators of dimension four, with renormalization conditions
\bey
{\cal O}^{\L=0\L_0(l)}_{k,l'~2}(0)=0~~;~~
\partial_{p^2}{\cal O}^{\L_R\L_0(l)}_{k,l'~2}|_{p=0}=\delta^{l,l'}~~;~~
{\cal O}^{\L_R\L_0(l)}_{k,l'~4}|_{p_i=0}=0\nonumber
\eny
and
\bey
{\cal O}^{\L=0\L_0(l)}_{g,l'~2}(0)=0~~;~
\partial_{p^2} {\cal O}_{g,l'~2}^{\L_R\L_0(l)}|_{p=0}=0~~;~~
{\cal O}^{\L_R\L_0(l)}_{g,l'~4}|_{p_i=0}=\delta^{l,l'}\nonumber
\eny
Using eq.(\ref{fluxO}) one could prove easily that
${\cal O}_{k,l'}^{\L\L_0(l)}[\Phi]={\cal
O}_{g,l'}^{\L\L_0(l)}[\Phi]=0$
for $l<l'$, since they are operators insertions with vanishing
renormalization conditions for $l<l'$.
Therefore we can define, as a formal series in $\hbar$, the functionals
$\hbar^{-l'}{\cal O}_{k,l'}^{\L\L_0}[\Phi]$ and
$\hbar^{-l'}{\cal O}_{g,l'}^{\L\L_0}[\Phi]$,
which again satisfy eq.(\ref{fluxO}).
The functionals 
${\cal O}_{g,0}^{\L\L_0}[\Phi]-\hbar^{-l'}{\cal O}_{g,l'}^{\L\L_0}[\Phi]$
have vanishing renormalization conditions and therefore are vanishing;
then it follows that
\bey
{\partial \over \partial \rho_3^{(0)}} L^{\L\L_0(l-l')}=
{\partial \over \partial \rho_3^{(l')}} L^{\L\L_0(l)} \nonumber
\eny
Using this fact we get
\bey
\sum_{l'=0}^l \beta_3^{(l')}{\partial \over \partial  \rho_3^{(l')} }
L^{\L\L_0(l)}[\L_R,\underline{\rho}_2,\underline{\rho}_3;\Phi]=
\sum_{l'=0}^l \beta_3^{(l')}{\partial \over \partial  \rho_3^{(0)} } 
 L^{\L\L_0(l-l')}[\L_R,\underline{\rho}_2,\underline{\rho}_3;\Phi] \nonumber
\eny
which can now be interpreted as the $l$-th term of the product of the series
$\beta_3$ and ${\partial \over \partial \rho^{(0)}_3} L^{\L\L_0}[\Phi]$. 
Similar considerations can be performed on the second term of
eq.(\ref{ERGG}).
Therefore one has
\bey
 \sum_{l'=1}^l \beta_2^{(l')} 
{\partial \over \partial \rho_2^{(l')} }
L^{\L\L_0(l)}[\L_R,\underline{\rho}_2,\underline{\rho}_3;\Phi]= 
 \sum_{l'=1}^l \beta_2^{(l')} 
{\partial \over \partial \rho_2^{(1)} }
L^{\L\L_0(l-l'+1)}[\L_R,\underline{\rho}_2,\underline{\rho}_3;\Phi] \nonumber
\eny
and this quantity is the $l$-th term of the product of the series
$\beta_2$ and 
$\hbar^{-1}{\partial \over  \partial \rho_2^{(1)}} L^{\L\L_0}[\Phi]$. 
Considering the case in which the only non-vanishing parameters are 
$\rho_3^{(0)}\equiv g$ and $\rho_2^{(1)}\equiv r$; 
as a consequence of the above equations, we get
\bea\label{ERGGG}
[\L_R {\partial \over \partial \L_R }
+\hbar^{-1} \beta_2 {\partial \over \partial r }
 + \beta_3{\partial \over \partial g } ] L^{\L\L_0}[\L_R,\hbar
r,g;\Phi] = 0
\ena
In the case $r=0$ we obtain eq.(\ref{ERG}).
Observe that we never introduced a parameter
$\rho_2^{(0)}$ for technical reasons.

Integrating formally eq.(\ref{ERGGG}) one arrives at the integral form
(\ref{ER}), where $\rho_2^{\L_R'\L_0}$ and  $\rho_3^{\L_R'\L_0}$
depend on $r$ and $g$ through the equations
\bey
\L_R'{d \over d\L_R'}\rho_2 = \beta_2(\rho_2,\rho_3)~~;~~
\L_R'{d \over d\L_R'}\rho_3 = \beta_3(\rho_2,\rho_3)\nonumber
\eny
with initial conditions
$\rho_2^{\L_R\L_0}=r\hbar$ and $\rho_3^{\L_R\L_0}=g$.

Let us finally make a comment on the statement made in the paper, that
the composite operators introduced in proving the renormalization
group equation are operators of dimension four which satisfy the
crucial equation (\ref{O22}); actually this is the only requirement
which is not trivially satisfied.
In all the cases considered, starting from the definition of the
functionals of the operators in term of the flow $L^{\L\L_0}[\Phi]$
it is an exercise with the usual induction scheme to prove the bound
(\ref{bs}) for $k=n-1$ and then eq.(\ref{O22}).


\begin{thebibliography}{99}



\bibitem{Rm} M. Gell-Mann and F.E. Low, Phys. Rev. {\bf 95} (1954) 1300.
\bibitem{Sy} K. Symanzik, Comm. Math. Phys. {\bf 16} (1970) 48; 
{\bf 23} (1971) 49.
\bibitem{Wil} K.G. Wilson and J.G. Kogut, Phys. Rep. {\bf C12} (1974) 75.
\bibitem{LM}J.H. Lowenstein and P.K. Mitter, Ann. Phys. {\bf 105} (1977) 138.
\bibitem{Qap} J.H. Lowenstein, Comm. Math. Phys. {\bf 24} (1971) 1; \\
Y.M.P. Lam, Phys. Rev. {\bf D6} (1971) 2145 ; {\bf D7} (1973) 2943.
\bibitem{PRR} M. Pernici, M. Raciti and F. Riva, `Hard-soft
renormalization and the exact renormalization group' preprint hep-th
9710145, to be published on Nucl. Phys. B.
\bibitem{wein} S. Weinberg, in Proceedings of the 1976 Int. School of
Subnuclear Physics, Erice.
\bibitem{Polchi}J. Polchinski, Nucl. Phys. {\bf B231} (1984) 269.
\bibitem{HL} J. Hughes and J. Liu, Nucl. Phys. {\bf B307} (1988) 183.
\bibitem{Gross} D. Gross, in `Methods in field theory' 1974 Les
Houches Lectures, ed. R. Balian and J. Zinn-Justin (North-Holland 1976).
\bibitem{Becchi2}C.M. Becchi, On the construction of renormalized quantum field
theory using renormalization group techniques,
in {\it Elementary particles, Field theory and Statistical
mechanics}, Eds. M. Bonini, G. Marchesini and E. Onofri,
Parma University 1993.
\bibitem{wein2} S. Weinberg, Phys. Rev. {\bf D8} (1973) 3497.
\bibitem{CS} C.G. Callan, Phys. Rev. {\bf D2} (1970) 1541;\\
K. Symanzik, Comm. Math. Phys. {\bf 18} (1970) 227. 
\bibitem{GS} M. Gomes and B. Schroer, Phys. Rev.  {\bf D10} (1974) 3525.
\bibitem{El} U. Ellwanger, Phys. Lett. {\bf B335} (1994) 364 ; 
Z. Phys. {\bf C76} (1997) 721.
\bibitem{ZJ} J. Zinn-Justin, 
{\it Quantum Field Theory and Critical Phenomena},\\
Oxford University Press 1989.
\bibitem{KKS} G. Keller, C. Kopper and M. Salmhofer, 
Helv. Phys. Acta {\bf 65} (1992) 32. 
\bibitem{BT} R.D. Ball and R.S. Thorne, 
Ann. Phys. {\bf 236} (1994) 117; Ann. Phys. {\bf 241} (1995). 
\bibitem{KKop} G. Keller and C. Kopper, Comm. Math. Phys. {\bf 148} (1992) 445.
\bibitem{BZ} E. Brezin, J.C. Le Guillou and J. Zinn-Justin,
`Field Theoretical Approach to Critical Phenomena' in
`Phase Transitions and Critical Phenomena' Vol.6, Ed. C. Domb and
M.S. Green (1976).
\bibitem{BDM}  M. Bonini, M. D'Attanasio and G. Marchesini, 
Nucl. Phys.  {\bf B409} (1993) 441.
\bibitem{KK} G. Keller and C. Kopper, Comm. Math. Phys. {\bf 161} (1994) 515.
\bibitem{Zimm} W. Zimmermann, Ann. Phys. {\bf 77} (1973) 536; Local
Operator Products and Renormalization,
in {\it Lectures on Elementary Particles and
Quantum Field Theory},  Eds. S. Deser, M. Grisaru and H. Pendleton, MIT
Press 1970.        
\end{thebibliography}
\end{document}